\documentclass[aps,pre,floatfix, showpacs,onecolumn,preprint]{revtex4-1}
\usepackage[colorlinks,linkcolor=blue,citecolor=blue,breaklinks]{hyperref}
\usepackage{natbib}
\usepackage{amssymb,amsfonts,amsmath}
\usepackage{graphicx}
\usepackage{units}
\usepackage{enumerate}
\usepackage{color}
\usepackage{placeins}
\usepackage[normalem]{ulem}
\usepackage[margin=3cm]{geometry}

\newcommand{\corr}{\mathcal{K}}

\newcommand{\cando}[1]{}

\DeclareFontFamily{U}{wncy}{}
    \DeclareFontShape{U}{wncy}{m}{n}{<->wncyr10}{}
    \DeclareSymbolFont{mcy}{U}{wncy}{m}{n}
    \DeclareMathSymbol{\sha}{\mathord}{mcy}{"58} 

\usepackage{array}
\usepackage{natbib}
\usepackage[normalem]{ulem}
 \usepackage{color}
 \definecolor{darkgreen}{rgb}{0,0.6,0}
 \definecolor{orange}{rgb}{0.99,0.257,0}

\usepackage{units}
\newlength{\FIGSIZE}  \newlength{\figsize}
\newcommand{\be}{\begin{align}} \newcommand{\ee}{\end{equation}} 
\newcommand{\ba}{\begin{eqnarray}} \newcommand{\ea}{\end{eqnarray}}
\newcommand{\eps}{\varepsilon}

\newcommand{\intl}{\int\limits}
\newcommand{\suml}{\sum\limits}

\newcommand{\e}[1]{eq.~(\ref{eq:#1})}

\newcommand{\es}[2]{eqs.~(\ref{eq:#1}, \ref{eq:#2})}

\newcommand{\E}[1]{Eq.~(\ref{eq:#1})}

\newcommand{\fig}[1]{Fig.~\ref{fig:#1}}

\newcommand{\Sec}[1]{Sec.~\ref{sec:#1}}

\newcommand{\app}[1]{Appendix~\ref{app:#1}}
\newcommand{\lrk}[1]{\left\langle #1 \right\rangle}

\newcommand{\s}{\sigma}

\newcommand{\til}{\widetilde}
\renewcommand{\hat}{\widehat}

\newcommand{\tr}{\tau_{\rm ref}}

\newcommand{\tm}{\til{m}}
\newcommand{\tJ}{\til{J}}
\newcommand{\tDelta}{\til{\Delta}}

\newcommand{\hyper}[1]{{}_2 F_1 \left( #1 \right) }

\allowdisplaybreaks[1]

\newcommand{\citein}[2][XXX]{#1 in ref \citep{#2}}

\allowdisplaybreaks[1]

\begin{document}

\title{Exact results for power spectrum and susceptibility of a leaky integrate-and-fire neuron with two-state noise}
\author{Felix Droste}
\email[]{fedro@physik.hu-berlin.de}
\author{Benjamin Lindner}

\affiliation{Bernstein Center for Computational Neuroscience, Haus 2, Philippstr 13, 10115 Berlin, Germany}
\affiliation{Department of Physics, Humboldt Universit\"at zu Berlin, Newtonstr 15, 12489 Berlin, Germany}

\date{\today}

\begin{abstract}
 The response properties of excitable systems driven by colored noise are of great interest, but are usually mathematically only accessible via approximations. For this reason, dichotomous noise, a rare example of a colored noise leading often to analytically tractable problems, has been extensively used in the study of stochastic systems.
 Here, we calculate exact expressions for the power spectrum and the susceptibility of a leaky integrate-and-fire neuron driven by asymmetric dichotomous noise.
 While our results are in excellent agreement with simulations, they also highlight a limitation of using dichotomous noise as a simple model for more complex fluctuations: Both power spectrum and susceptibility exhibit an undamped periodic structure, the origin of which we discuss in detail.
\end{abstract}

\maketitle

\section{Introduction}

An important class of non-equilibrium systems are excitable systems \citep{LinGar04}, in which small perturbations can lead to large excursions.
Examples include lasers \citep{GiuGre97}, chemical reactions \citep{Mer92}, or neurons \citep{Izh07}, in which the excitation corresponds to the emission of an action potential. 
Both the spontaneous fluctuations of such a system (characterized, for instance, by its power spectrum) as well as its response to a time-dependent external driving (quantified for weak signals by the susceptibility or transfer function) are of great interest.
Often, a realistic description requires incorporating also the correlation structure of the noise that the system is subject to; this means that the popular assumption of a Gaussian white noise does not always hold and one has to deal with a colored, potentially non-Gaussian noise \citep{HanJun95}. 

In the stochastic description of neurons, power spectrum and susceptibility are of particular interest because they are closely linked to measures of information transmission \citep{BorThe99}. 
An important model class are stochastic integrate-and-fire (IF) neurons \citep{Bur06a,Bur06b}, the response properties of which have received considerable attention over the last decades.
Exact results for the susceptibility have been derived for leaky IF neurons (LIF) driven by Gaussian white noise \citep{BruCha01, LinLSG01} or white shot-noise with exponentially distributed weights \citep{RicSwa10}. 
For IF neurons driven by exponentially correlated Gaussian noise, only approximate results in the limit of high frequencies \citep{FouBru02, FouHan03} and short  \citep{AliRic11, SchDie15} or long \citep{AliRic11} noise correlation time exist. 
The power spectrum is exactly known for perfect IF (PIF) neurons \citep{SteFre72} and LIF neurons driven by white Gaussian noise \citep{LinLSG02}. 
For colored noise, approximate results for the auto-correlation function (the Fourier transform of the power spectrum) exist for LIF neurons in the limit of long noise correlation time \citep{MorPar06} and for PIF neurons driven by weak, arbitrarily colored noise \citep{SchDro15}.

The dichotomous Markov process (DMP) \citep{HorLef84, Ben06}, a two-state noise with exponential correlation function, is the rare example of a driving colored noise that can lead to tractable problems.
For this reason, it has been extensively used in the statistical physics literature for a long time \citep{HorLef84, HanJun95}; recently, its use as a model of neural input has been growing \citep{SalSej02, Lin04, DroLin14, MulDro15, ManLum16}.
Known exact results for IF neurons driven by a DMP include the firing rate and coefficient of variation (CV) of PIF and LIF \citep{SalSej02} or arbitrary IF neurons \citep{DroLin14}, the interspike-interval (ISI) density and serial correlation coefficients (SCC) of ISIs for PIF \citep{Lin04, MulDro15} and LIF neurons \citep{ManLum16}, the stationary voltage distribution for arbitrary IF neurons \citep{DroLin14}, or the power spectrum for PIF neurons \citep{MulDro15}.

In this work, we consider an LIF neuron driven by asymmetric dichotomous noise and calculate exact expressions for the spontaneous power spectrum and the susceptibility, i.e. the rate response to a signal that is modulating the additive drive to the neuron.
The outline is as follows. We briefly present the model and describe the associated master equation in \Sec{model}.
In \Sec{powspec}, we derive an expression for the power spectrum and discuss its peculiar structure. 
Here, our approach was inspired by a numerical scheme for white-noise-driven IF neurons \citep{Ric08}.
Reusing results for the power spectrum, we calculate the susceptibility in \Sec{suscep}, employing a perturbation ansatz similar to approaches previously used for Gaussian noise \citep{FouBru02}.
In \Sec{broadband}, we study numerically how robust our results are when using broadband signals.
We close with a short summary and some concluding remarks in \Sec{summary}.

\section{Model and master equation}

\label{sec:model}

\begin{figure*}
 \includegraphics[scale=1]{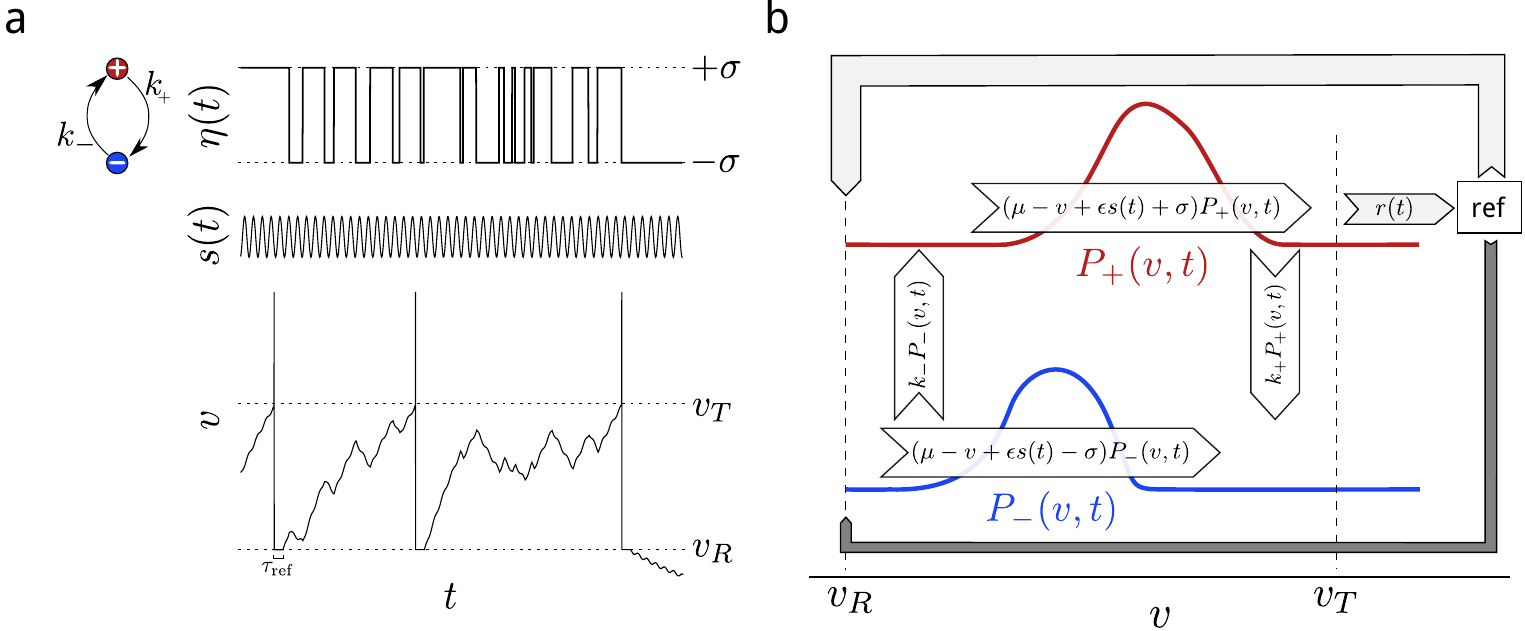}
 \caption{\label{fig:scheme} {\bf An LIF neuron driven by dichotomous noise and a weak signal} {\bf (a)} A dichotomous noise $\eta(t)$ jumps between a plus state and a minus state at rates $k_+$ and $k_-$. Shown is a sample realization of $\eta(t)$ and the signal $s(t)$ (here a sinusoid) along with the resulting trajectory of the voltage $v$. The spikes at threshold crossings are not dynamically generated but added for the purpose of illustration. {\bf (b)} Sketch of the probability densities and fluxes in the master equation.  }
\end{figure*}

The evolution of the membrane potential $v$ of an LIF neuron is governed by
\begin{equation}
 \dot{v} = \mu - v + \eps s(t) + \eta(t). \label{eq:dynamics_lif}
\end{equation}
Spiking is implemented through an explicit fire-and-reset rule: when the voltage hits a threshold $v_T$, it is reset to $v_R$, where it remains clamped for a refractory period $\tr$.
In \e{dynamics_lif}, $\mu$ sets the equilibrium potential, $\eps s(t)$ is a weak stimulus and $\eta(t)$ is a potentially asymmetric Markovian dichotomous noise. 
Time is measured in units of the membrane time constant.

The dichotomous noise $\eta(t)$ jumps between the two values $\s$ and $-\s$ at the constant rates $k_+$ (jumping from the "plus state" $\s$ to the "minus state" $-\s$) and $k_-$ (jumping from $-\s$ to $\s$; see \fig{scheme}a).
Note that this can always be transformed to a noise with asymmetric noise values and an additional offset.
The properties of such a process are rather straightforward to calculate and have been known for a long time \citep{Fit83, HorLef84, Gar85}.
In the following, we will need the transition probabilities, i.e. the probability to find the noise in state $i$ given that it was in state $j$ a certain time $\tau$ before,
\begin{equation}
 P_{i|j}(\tau) := \Pr\left(\eta(t+\tau)=\s_i|\eta(t)= \s_j\right),
\end{equation}
where $\{i,j\} = \{+,-\}, \s_+ = \s$, $\s_- = -\s$.
We will only need the transition probabilities conditioned on starting in the plus state, which read \citep{Gar85}
\begin{align}
  P_{+|+}(\tau) &= \frac{k_+ e^{-(k_+ + k_-) \tau} + k_-}{k_+ + k_-}, \label{eq:dicho_Ppp}\\
  P_{-|+}(\tau) &= \frac{k_+}{k_+ + k_-} \left( 1 - e^{-(k_+ + k_-) \tau} \right). \label{eq:dicho_Ppm}\\
\end{align}

In this paper, we limit ourselves to the case $\mu - \s < v_T$. 
This means that the neuron can only cross the threshold when the noise is in the plus state. 
A general treatment for different parameter regimes, as has been carried out for stationary density and first-passage-time moments in ref. \citep{DroLin14}, involves much book-keeping and is beyond the scope of this work.
Note that this choice of parameters does not constrain the neuron to a fluctuation-driven (sub-threshold) or mean-driven (supra-threshold) regime, both $\mu + \lrk{\eta(t)} > v_T$ and $\mu + \lrk{\eta(t)} < v_T$ are still possible. Our choice, however, implies that the generated spike train in the absence of a signal ($\varepsilon=0$) is a renewal process: Because firing occurs only in the plus state and the noise has no memory about the past (Markov property), the interspike intervals are statistically independent.

A common approach to the description of systems driven by dichotomous noise is to consider two probabilities: $P_+(v,t) dv$, the probability to find the noise in the plus state and the voltage in the interval $(v, v+dv)$ at time $t$, and, analogously, $P_-(v,t) dv$ (see the scheme in \fig{scheme}b).
The system is then described by the following master equation, 
\begin{align}
 \begin{split}
 \partial_t P_+(v, t) &= - \partial_v \left[ (\mu-v + \eps s(t) +\s) P_+(v, t) \right]\\
 &\quad - k_+ P_+(v, t) + k_- P_-(v, t) \\
 &\quad + r(t-\tr) P_{+|+}(\tr) \delta(v-v_R) - r(t) \delta(v-v_T),
 \end{split} \label{eq:modulated_mastereq_pp} \\
 \begin{split} 
 \partial_t P_-(v, t) &= - \partial_v \left( (\mu-v+\eps s(t) -\s) P_-(v, t) \right) 
\\
  &\quad+ k_+ P_+(v, t) - k_- P_-(v, t) \\
  &\quad+ r(t-\tr) P_{-|+}(\tr) \delta(v-v_R). \label{eq:modulated_mastereq_pm}
  \end{split}
\end{align}
The boundary conditions are $P_+(v_T^+) = 0$ and $P_-(v_T^+) = 0$.
If $\mu - \s > v_R$ (the minus dynamics has a stable fixed point between $v_R$ and $v_T$), one needs additionally $P_+(v_R^-) = 0$, $P_-(v_R^-) = 0$ (see \citep{DroLin14, Dro15} for a detailed treatment of fixed points in DMP-driven IF neurons). Here, $v_T^+$ ($v_R^-$) refers to a voltage infinitesimally above $v_T$ (below $v_R$).

The respective first two lines of \e{modulated_mastereq_pp} and \e{modulated_mastereq_pm} are similar to what one would have for other systems driven by dichotomous noise; they describe the deterministic drift within each state and the switching between states.
The third line is more particular to this neuronal setup and incorporates the fire-and-reset rule: trajectories are taken out at the threshold $v_T$ and the outflux corresponds to the instantaneous firing rate $r(t)$; after the refractory period $\tr$ has passed, they are reinserted at $v_R$. 
As we assume $\mu - \s < v_T$, trajectories can only leave the system in the plus state; however, they can get reinserted in both states because the noise may have changed its state during the refractory period. 
This is captured by the transition probabilities $P_{\pm|+}(\tr)$.
Note that one can describe the same dynamics by omitting these source and sink terms (the $\delta$ function inhomogeneities) and instead using more complicated boundary conditions, $P_+(v_T) = r(t)/(\mu-v_T+\eps s(t) + \s), P_-(v_T) = 0$, and jump conditions at $v_R$: $[P_+(v)]_{v_R} := \lim_{\delta \to 0} P_+(v_R + \delta) - P_+(v_R - \delta) = r(t-\tr) P_{+|+}(\tr) / (\mu-v_R+\eps s(t) + \s)$ and  $[P_-(v)]_{v_R} = r(t-\tr) P_{-|+}(\tr) / (\mu-v_R+\eps s(t) + \s)$.
If $\mu - \s > v_R$, one additionally needs $P_+(v_R^-) = 0$ and $P_-(v_R^-) = 0$. 

\section{Power spectrum}
\label{sec:powspec}

For the calculation of the spontaneous power spectrum, we set $\eps=0$ in \e{dynamics_lif} and \es{modulated_mastereq_pp}{modulated_mastereq_pm}.
According to the Wiener-Khinchin theorem \citep{Gar85,Ris89}, the power spectrum is the Fourier transform of the auto-correlation function of the spike train,
\begin{equation}
 S(f) = \intl_{-\infty}^{\infty} d\tau\; e^{2\pi i f \tau} \corr(\tau).
\end{equation}
The auto-correlation function can be expressed in terms of the stationary rate $r_0$ and the spike-triggered rate $m(\tau)$,
\begin{equation}
 \corr(\tau) = r_0 m(\tau) + r_0 \delta(\tau) - r_0^2.
\end{equation}
For the case considered here, the stationary rate reads \citep{DroLin14, Dro15}
\begin{equation}
 \begin{split}
 r_0 &= \left[ \tr + (k_+ + k_-) \intl_{v_R}^{v_T}dx\; \intl_{x}^{\mu-\s}dy\; 
  \frac{ \left| \frac{\mu-y+\s}{\mu-x+\s} \right|^{k_+} \left| \frac{\mu-y-\s}{\mu-x-\s} \right|^{k_-}  }{(\mu-x+\s)(\mu-y-\s)} \right. \\
  &\quad+ \left. \frac{1-e^{-\tr(k_+ + k_-)}}{k_+ + k_-} \left( -1 + (k_+ + k_-) \intl_{v_R}^{\mu-\s} dx\; \frac{ \left| \frac{\mu-x+\s}{\mu-v_R+\s} \right|^{k_+} \left| \frac{\mu-x-\s}{\mu-v_R-\s} \right|^{k_-}  }{\mu-x-\s} \right) \right]^{-1}.
 \end{split}
\end{equation}
The spike-triggered rate is the rate at which spikes occur at time $t=\tau$ given that there was a (different) spike at $t=0$.
The power spectrum can be expressed using the Fourier transform of the spike-triggered rate, $\til{m}(f)$,
\begin{equation}
 \label{eq:powspec_via_strate}
  S(f) = r_0 \left(1 + 2 \Re[\til{m}(f)] \right).
\end{equation}

To calculate $\til{m}(f)$, we modify the master equation \es{modulated_mastereq_pp}{modulated_mastereq_pm},
\begin{align}
 \begin{split}
 \partial_t P_+(v, t) &= - \partial_v \left( (\mu-v+\s) P_+(v, t) \right) - k_+ P_+(v, t) + k_- P_-(v, t) \\
 &\quad + m(t-\tr) P_{+|+}(\tr) \delta(v-v_R) - m(t) \delta(v-v_T) \\
 &\quad + \delta(t-\tr) P_{+|+}(\tr) \delta(v-v_R),
 \end{split} \label{eq:strate_master1} \\
 \begin{split} 
 \partial_t P_-(v, t) &= - \partial_v \left( (\mu-v-\s) P_-(v, t) \right) + k_+ P_+(v, t) - k_- P_-(v, t) \\
  &\quad+ m(t-\tr) P_{-|+}(\tr) \delta(v-v_R) \\
  &\quad+ \delta(t-\tr) P_{-|+}(\tr) \delta(v-v_R),
 \label{eq:strate_master2}
 \end{split}
\end{align}
with the boundary conditions $P_+(v_T^+, t) = 0, P_-(v_T^+, t) = 0$ and, if $\mu-\s>v_R$, also $P_+(v_R^-, t) = 0, P_-(v_R^-, t) = 0$. Further, the initial condition is $P_+(v, 0^-) = P_-(v, 0^-) = 0$.

In \e{strate_master1}, the source and sink terms $m(t-\tr) P_{+|+}(\tr) \delta(v-v_R) - m(t) \delta(v-v_T)$ implement the fire-and-reset rule (trajectories cross the threshold at a rate $m(t)$ and get inserted at $v_R$ after the refractory period $\tr$ has passed). 
The term $\delta(t-\tr) P_{+|+}(\tr)$, together with the initial condition, accounts for the fact that the neuron has fired at $t=0$, so that after the refractory period, all probability starts at $v_R$ (a fraction $P_{+|+}(\tr)$ in the plus state). 
Equivalent considerations apply to \e{strate_master2}.

The system of two first-order partial differential equations for the probability density can be transformed to a ordinary, second-order differential equation for the Fourier transform of the probability flux, $\tJ(v,f) = \int_{-\infty}^{\infty} dt\; e^{2\pi i f t} J(v,t)$, where $J(v,t) = J_+(v,t) + J_-(v,t) = (v-\mu+\s)P_+(v,t) + (v-\mu-\s)P_-(v,t)$.
After some simplifying steps, it reads:
\begin{align}
 \begin{split}
 \label{eq:j_ode}
   0 &= \tJ''(z) + p(z) \tJ'(z) + q(z) \tJ(z) \\
   &\quad - \left( p(z) + \frac{2\pi i f}{1-z} \right) 2 \s \tDelta_+(z) - \left( p(z) - \frac{2\pi i f}{z} \right) 2 \s \tDelta_-(z) \\
   &\quad - 2\s  \big[ \tDelta_+'(z) + \tDelta_-'(z) \big],
 \end{split}
\end{align}
with the boundary conditions $J(z_T^+) = 0, J'(z_T^+) = 0; J(\min[z_R^-, 0]) = 0, J'(\min[z_R^-,0]) = 0$.
Here, we have omitted the $f$ argument for the sake of readability, have made the change of variables
\begin{equation}
 z := \frac{v - \mu + \s}{2 \s},
\end{equation}
and used the abbreviations
\begin{align}
 p(z) &= \frac{-z \left( 2 - k_+ - k_- + 4\pi i f \right) + \left( 1 - k_- + 2\pi i f \right) }{z(1-z)}, \\
 q(z) &= \frac{-2\pi i f \left( 1 - k_+ - k_-+ 2\pi i f \right)}{z(1-z)}, \\
  \tDelta_+(z) &= \tm(f) \frac{1}{2\s} \big[ e^{2\pi i f\tr} P_{+|+}(\tr) \delta(z-z_R) - \delta(z-z_T) \big] + \frac{1}{2\s}  e^{2\pi i f\tr} P_{+|+}(\tr) \delta(z-z_R), \label{eq:Deltap} \\
  \tDelta_-(z) &= \tm(f) \frac{1}{2\s} e^{2\pi i f\tr} P_{-|+}(\tr) \delta(z-z_R) + \frac{1}{2\s}  e^{2\pi i f\tr} P_{-|+}(\tr) \delta(z-z_R).\label{eq:Deltam} 
\end{align}
The homogeneous part of \e{j_ode} can be identified with the hypergeometric differential equation \citep{MorFes53}, for which solutions are known.
By constructing a solution to the inhomogeneous ODE that fulfills the boundary conditions, we show in \app{solsimpl} that
\begin{equation}
 \label{eq:jzerosol}
 0 = \intl_{-\infty}^{\infty} du\; (k_- - 2\pi i f) \tDelta_+(u) \mathcal{F}(u, f) + k_- \tDelta_-(u) \mathcal{G}(u, f),
\end{equation}
for general inhomogeneities $\tDelta_\pm(u)$, provided that they vanish outside $[\min(z_R,0), z_T]$. 
Here, $\mathcal{F}(u, f)$ and $\mathcal{G}(u, f)$ are given in terms of hypergeometric functions \citep{AbrSte72},
\begin{align}
 \mathcal{F}(z,f)  &:= \hyper{-2\pi i f,k_+ + k_- - 2\pi i f;k_- - 2\pi i f;z}, \label{eq:fdef} \\
 \mathcal{G}(z, f) &:= \hyper{-2\pi i f,k_+ + k_- - 2\pi i f;1+k_- - 2\pi i f;z} \label{eq:gdef}.
\end{align}
\E{jzerosol} can then be solved for the spike-triggered rate, which is contained in $\tDelta_\pm(u)$.
Owing to the delta functions in the inhomogeneities, the integration in \e{jzerosol} is straightforward to carry out. 
For the power spectrum, one obtains, via \e{powspec_via_strate},
\begin{equation}
\begin{split}
 \label{eq:powspec}
 S(f) &= r_0 \frac{\left| e^{-2\pi i f\tr} \mathcal{F}(z_T,f) \right|^2-\left| P_{+|+}(\tr) \mathcal{F}(z_R,f) + \frac{k_-}{k_- - 2\pi i f} P_{-|+}(\tr) \mathcal{G}(z_R,f) \right|^2}{\left| e^{-2\pi i f\tr} \mathcal{F}(z_T,f) -  P_{+|+}(\tr) \mathcal{F}(z_R,f) - \frac{k_-}{k_- - 2\pi i f} P_{-|+}(\tr) \mathcal{G}(z_R,f)\right|^2}.
\end{split}
\end{equation}
This is the first central result of this work.

For the special case of a vanishing refractory period, $\tr = 0$, \e{powspec} takes a particularly compact form
\begin{align}
 \label{eq:powspec_tr0}
 S(f) 
        &= r_0 \frac{\left| \mathcal{F}(z_T, f) \right|^2 - \left| \mathcal{F}(z_R, f) \right|^2}{\left|\mathcal{F}(z_T, f) - \mathcal{F}(z_R, f) \right|^2}.
\end{align}
which resembles the form of the expression for the power spectrum of LIF neurons driven by Gaussian white noise \citep{LinLSG02}.

\begin{figure*}
 \newcommand{\paramvalue}[2][]{\protect %
 \ifnum\pdfstrcmp{#2}{vhist_r}=0 \ifnum\pdfstrcmp{#1}{dec}=0 1.1 \else\ifnum\pdfstrcmp{#1}{sci}=0 1.1 \times 10^{0} \else\ifnum\pdfstrcmp{#1}{dec0}=0 1 \else\ifnum\pdfstrcmp{#1}{dec1}=0 1.1 \else\ifnum\pdfstrcmp{#1}{dec2}=0 1.1 \else 1.1\fi\fi\fi\fi\fi \else %
 \ifnum\pdfstrcmp{#2}{vr}=0 \ifnum\pdfstrcmp{#1}{dec}=0 0 \else\ifnum\pdfstrcmp{#1}{sci}=0 0 \else\ifnum\pdfstrcmp{#1}{dec0}=0 0 \else\ifnum\pdfstrcmp{#1}{dec1}=0 0 \else\ifnum\pdfstrcmp{#1}{dec2}=0 0 \else 0\fi\fi\fi\fi\fi \else %
 \ifnum\pdfstrcmp{#2}{seed}=0 0 \else %
 \ifnum\pdfstrcmp{#2}{vt}=0 \ifnum\pdfstrcmp{#1}{dec}=0 1 \else\ifnum\pdfstrcmp{#1}{sci}=0 1 \times 10^{0} \else\ifnum\pdfstrcmp{#1}{dec0}=0 1 \else\ifnum\pdfstrcmp{#1}{dec1}=0 1 \else\ifnum\pdfstrcmp{#1}{dec2}=0 1 \else 1\fi\fi\fi\fi\fi \else %
 \ifnum\pdfstrcmp{#2}{vhist_l}=0 \ifnum\pdfstrcmp{#1}{dec}=0 -0.1 \else\ifnum\pdfstrcmp{#1}{sci}=0 -1 \times 10^{-1} \else\ifnum\pdfstrcmp{#1}{dec0}=0 -0 \else\ifnum\pdfstrcmp{#1}{dec1}=0 -0.1 \else\ifnum\pdfstrcmp{#1}{dec2}=0 -0.1 \else -0.1\fi\fi\fi\fi\fi \else %
 \ifnum\pdfstrcmp{#2}{integ_infty}=0 \ifnum\pdfstrcmp{#1}{dec}=0 15 \else\ifnum\pdfstrcmp{#1}{sci}=0 1.5 \times 10^{1} \else\ifnum\pdfstrcmp{#1}{dec0}=0 15 \else\ifnum\pdfstrcmp{#1}{dec1}=0 15 \else\ifnum\pdfstrcmp{#1}{dec2}=0 15 \else 15\fi\fi\fi\fi\fi \else %
 \ifnum\pdfstrcmp{#2}{f_max}=0 \ifnum\pdfstrcmp{#1}{dec}=0 100 \else\ifnum\pdfstrcmp{#1}{sci}=0 1 \times 10^{2} \else\ifnum\pdfstrcmp{#1}{dec0}=0 100 \else\ifnum\pdfstrcmp{#1}{dec1}=0 100 \else\ifnum\pdfstrcmp{#1}{dec2}=0 100 \else 100\fi\fi\fi\fi\fi \else %
 \ifnum\pdfstrcmp{#2}{tr}=0 \ifnum\pdfstrcmp{#1}{dec}=0 0.1 \else\ifnum\pdfstrcmp{#1}{sci}=0 1 \times 10^{-1} \else\ifnum\pdfstrcmp{#1}{dec0}=0 0 \else\ifnum\pdfstrcmp{#1}{dec1}=0 0.1 \else\ifnum\pdfstrcmp{#1}{dec2}=0 0.1 \else 0.1\fi\fi\fi\fi\fi \else %
 \ifnum\pdfstrcmp{#2}{f_sig}=0 \ifnum\pdfstrcmp{#1}{dec}=0 -1 \else\ifnum\pdfstrcmp{#1}{sci}=0 -1 \times 10^{0} \else\ifnum\pdfstrcmp{#1}{dec0}=0 -1 \else\ifnum\pdfstrcmp{#1}{dec1}=0 -1 \else\ifnum\pdfstrcmp{#1}{dec2}=0 -1 \else -1\fi\fi\fi\fi\fi \else %
 \ifnum\pdfstrcmp{#2}{vhist_N}=0 100 \else %
 \ifnum\pdfstrcmp{#2}{eps_mu}=0 \ifnum\pdfstrcmp{#1}{dec}=0 0 \else\ifnum\pdfstrcmp{#1}{sci}=0 0 \else\ifnum\pdfstrcmp{#1}{dec0}=0 0 \else\ifnum\pdfstrcmp{#1}{dec1}=0 0 \else\ifnum\pdfstrcmp{#1}{dec2}=0 0 \else 0\fi\fi\fi\fi\fi \else %
 \ifnum\pdfstrcmp{#2}{N_trials}=0 10000 \else %
 \ifnum\pdfstrcmp{#2}{df}=0 \ifnum\pdfstrcmp{#1}{dec}=0 0.01 \else\ifnum\pdfstrcmp{#1}{sci}=0 1 \times 10^{-2} \else\ifnum\pdfstrcmp{#1}{dec0}=0 0 \else\ifnum\pdfstrcmp{#1}{dec1}=0 0 \else\ifnum\pdfstrcmp{#1}{dec2}=0 0.01 \else 0.01\fi\fi\fi\fi\fi \else %
 \ifnum\pdfstrcmp{#2}{eps_km}=0 \ifnum\pdfstrcmp{#1}{dec}=0 0 \else\ifnum\pdfstrcmp{#1}{sci}=0 0 \else\ifnum\pdfstrcmp{#1}{dec0}=0 0 \else\ifnum\pdfstrcmp{#1}{dec1}=0 0 \else\ifnum\pdfstrcmp{#1}{dec2}=0 0 \else 0\fi\fi\fi\fi\fi \else %
 \ifnum\pdfstrcmp{#2}{dt}=0 \ifnum\pdfstrcmp{#1}{dec}=0 0.001 \else\ifnum\pdfstrcmp{#1}{sci}=0 1 \times 10^{-3} \else\ifnum\pdfstrcmp{#1}{dec0}=0 0 \else\ifnum\pdfstrcmp{#1}{dec1}=0 0 \else\ifnum\pdfstrcmp{#1}{dec2}=0 0 \else 0.001\fi\fi\fi\fi\fi \else %
 \ifnum\pdfstrcmp{#2}{integ_epsabs}=0 \ifnum\pdfstrcmp{#1}{dec}=0 0.0001 \else\ifnum\pdfstrcmp{#1}{sci}=0 1 \times 10^{-4} \else\ifnum\pdfstrcmp{#1}{dec0}=0 0 \else\ifnum\pdfstrcmp{#1}{dec1}=0 0 \else\ifnum\pdfstrcmp{#1}{dec2}=0 0 \else 1 \times 10^{-4}\fi\fi\fi\fi\fi \else %
 \ifnum\pdfstrcmp{#2}{f_c}=0 \ifnum\pdfstrcmp{#1}{dec}=0 1 \else\ifnum\pdfstrcmp{#1}{sci}=0 1 \times 10^{0} \else\ifnum\pdfstrcmp{#1}{dec0}=0 1 \else\ifnum\pdfstrcmp{#1}{dec1}=0 1 \else\ifnum\pdfstrcmp{#1}{dec2}=0 1 \else 1\fi\fi\fi\fi\fi \else %
 \ifnum\pdfstrcmp{#2}{km}=0 [2.0, 20.0] \else %
 \ifnum\pdfstrcmp{#2}{mu}=0 \ifnum\pdfstrcmp{#1}{dec}=0 0.8 \else\ifnum\pdfstrcmp{#1}{sci}=0 8 \times 10^{-1} \else\ifnum\pdfstrcmp{#1}{dec0}=0 1 \else\ifnum\pdfstrcmp{#1}{dec1}=0 0.8 \else\ifnum\pdfstrcmp{#1}{dec2}=0 0.8 \else 0.8\fi\fi\fi\fi\fi \else %
 \ifnum\pdfstrcmp{#2}{s}=0 \ifnum\pdfstrcmp{#1}{dec}=0 2.4 \else\ifnum\pdfstrcmp{#1}{sci}=0 2.4 \times 10^{0} \else\ifnum\pdfstrcmp{#1}{dec0}=0 2 \else\ifnum\pdfstrcmp{#1}{dec1}=0 2.4 \else\ifnum\pdfstrcmp{#1}{dec2}=0 2.4 \else 2.4\fi\fi\fi\fi\fi \else %
 \ifnum\pdfstrcmp{#2}{kp}=0 eval_after_unroll(km/2) \else %
 \ifnum\pdfstrcmp{#2}{model}=0 lif \else %
 \PackageError{paramvalue}{unknown param name: #2}{} 
 \fi\fi\fi\fi\fi\fi\fi\fi\fi\fi\fi\fi\fi\fi\fi\fi\fi\fi\fi\fi\fi\fi}
 
 \includegraphics{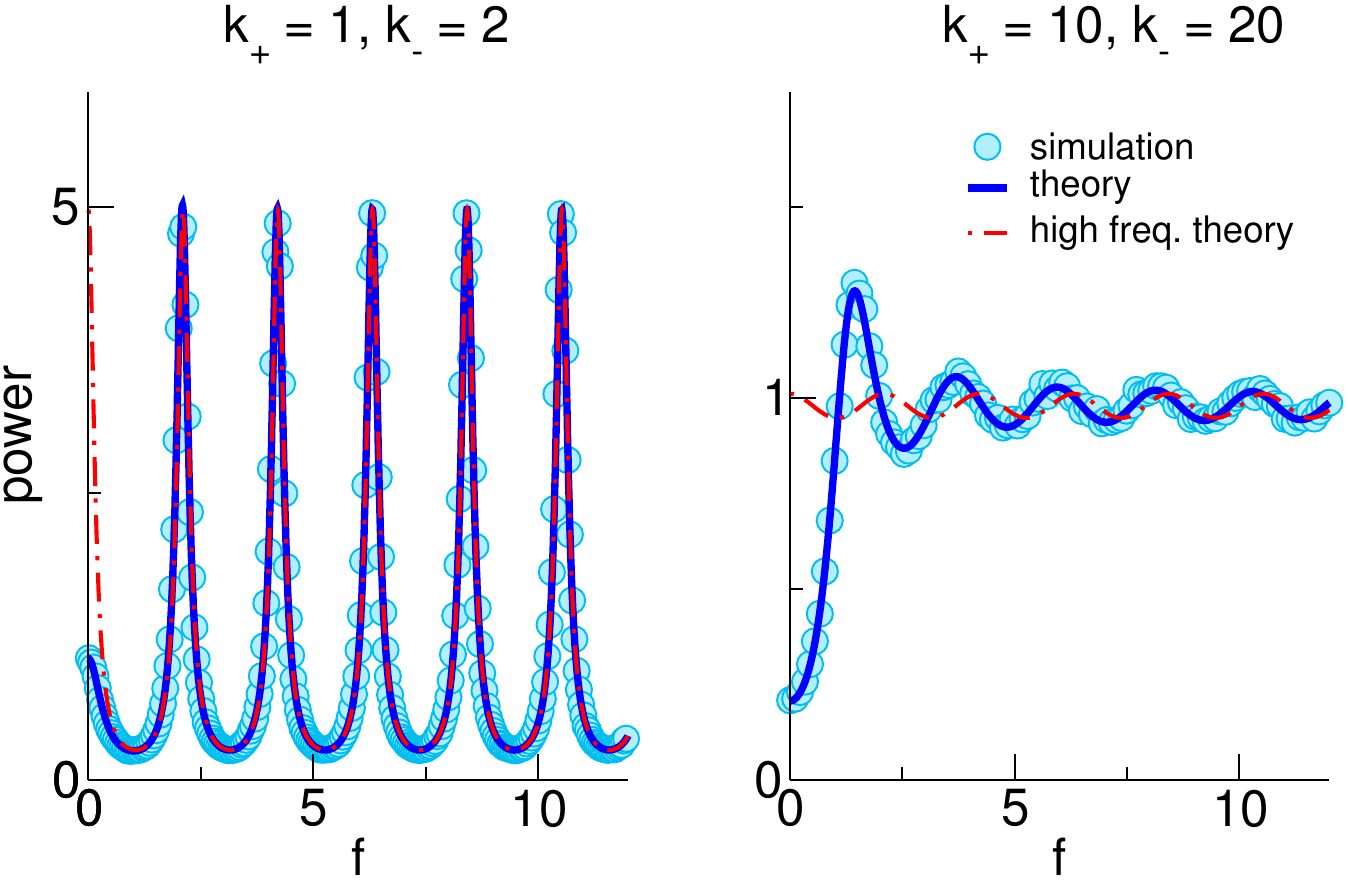}
 \caption{\label{fig:powspec} {\bf Power spectra for two different combinations of noise switching rates} Shown are simulation results (light blue solid lines), compared to the exact theoretical expression  (\e{powspec}, dark blue solid lines) and the high-frequency limit  (\e{powspec_highfreq}, red dash-dotted line). Remaining parameters: $\mu = \paramvalue{mu}, \tr=\paramvalue{tr}, \s=\paramvalue{s}, v_R=\paramvalue{vr}, v_T=\paramvalue{vt}$, $\eps=\paramvalue{eps_mu}$. }
\end{figure*}

In \fig{powspec}, we plot the power spectrum and compare it to simulations.
It is apparent that the theory is in excellent agreement with simulation results.
The most striking feature of the power spectrum, especially for slow switching of the noise, is an undamped oscillation.
This is in stark contrast to what one usually expects from spike train power spectra \citep{BaiKoc94, GabKoc98}, which saturate at the firing rate $r_0$.
This periodicity in the spectrum, which has been previously observed in the PIF model \citep{MulDro15}, can also be seen explicitly in the analytics by taking \e{powspec} to its high-frequency limit,
\begin{equation}
 \label{eq:powspec_highfreq}
 \begin{split}
 &S(f \gg 1) = 
 r_0 \frac{1-P_{+|+}^2(\tr)e^{-2k_+ (T_d^+ - \tr)}}{1+P_{+|+}^2(\tr)e^{-2k_+ (T_d^+ - \tr)}-2 P_{+|+}(\tr)e^{-k_+ (T_d^+ - \tr)} \cos(2\pi f T_d^+)},
 \end{split}
\end{equation}
where $T_d^+$ is the (deterministic) time from reset to threshold in the plus state,
\begin{equation}
 \label{eq:Tdp}
 T_d^+ = \ln\left( \frac{\mu+\s-v_R}{\mu+\s-v_T} \right) + \tr = \ln\left( \frac{1-z_R}{1-z_T} \right) + \tr.
\end{equation}
Taking the Gaussian-white-noise limit \citep{Bro83} $\s = \sqrt{2Dk}, k := k_+ = k_- \to \infty$ (where $D$ is the noise intensity of the resulting process) in \e{powspec_highfreq} yields $S(f \gg 1) \xrightarrow{k\to\infty} r_0$, which is the known high frequency behavior for the Gaussian white noise case. 

Introducing a modified switching rate,
\begin{equation}
 \label{eq:khat}
 \hat{k}_+ := k_+ \left(1-\frac{\tr}{T_d^+}\right) - \frac{\ln\left[P_{+|+}(\tr)\right]}{T_d^+},
\end{equation}
\e{powspec_highfreq} can be written compactly as 
\begin{equation}
\label{eq:powspec_highfreq_compact}
 S(f \gg 1) = r_0 \frac{\sinh\left(\hat{k}_+ T_d^+\right)}{\cosh\left(\hat{k}_+ T_d^+\right) - \cos\left(2\pi f T_d^+\right)}.
\end{equation}

The high-frequency limit is also shown in \fig{powspec}.
For slow switching, it is indistinguishable (within line thickness) from the exact theory over most of the shown frequency range and only deviates from it for small frequencies. 


\begin{figure*}
 \newcommand{\paramvalue}[2][]{\protect %
 \ifnum\pdfstrcmp{#2}{_N_trajs}=0 50 \else %
 \ifnum\pdfstrcmp{#2}{f_max}=0 \ifnum\pdfstrcmp{#1}{dec}=0 100 \else\ifnum\pdfstrcmp{#1}{sci}=0 1 \times 10^{2} \else\ifnum\pdfstrcmp{#1}{dec0}=0 100 \else\ifnum\pdfstrcmp{#1}{dec1}=0 100 \else\ifnum\pdfstrcmp{#1}{dec2}=0 100 \else 100\fi\fi\fi\fi\fi \else %
 \ifnum\pdfstrcmp{#2}{vhist_r}=0 \ifnum\pdfstrcmp{#1}{dec}=0 1.1 \else\ifnum\pdfstrcmp{#1}{sci}=0 1.1 \times 10^{0} \else\ifnum\pdfstrcmp{#1}{dec0}=0 1 \else\ifnum\pdfstrcmp{#1}{dec1}=0 1.1 \else\ifnum\pdfstrcmp{#1}{dec2}=0 1.1 \else 1.1\fi\fi\fi\fi\fi \else %
 \ifnum\pdfstrcmp{#2}{eps_mu}=0 \ifnum\pdfstrcmp{#1}{dec}=0 0 \else\ifnum\pdfstrcmp{#1}{sci}=0 0 \else\ifnum\pdfstrcmp{#1}{dec0}=0 0 \else\ifnum\pdfstrcmp{#1}{dec1}=0 0 \else\ifnum\pdfstrcmp{#1}{dec2}=0 0 \else 0\fi\fi\fi\fi\fi \else %
 \ifnum\pdfstrcmp{#2}{N_trials}=0 1000 \else %
 \ifnum\pdfstrcmp{#2}{df}=0 \ifnum\pdfstrcmp{#1}{dec}=0 0.01 \else\ifnum\pdfstrcmp{#1}{sci}=0 1 \times 10^{-2} \else\ifnum\pdfstrcmp{#1}{dec0}=0 0 \else\ifnum\pdfstrcmp{#1}{dec1}=0 0 \else\ifnum\pdfstrcmp{#1}{dec2}=0 0.01 \else 0.01\fi\fi\fi\fi\fi \else %
 \ifnum\pdfstrcmp{#2}{vr}=0 \ifnum\pdfstrcmp{#1}{dec}=0 0 \else\ifnum\pdfstrcmp{#1}{sci}=0 0 \else\ifnum\pdfstrcmp{#1}{dec0}=0 0 \else\ifnum\pdfstrcmp{#1}{dec1}=0 0 \else\ifnum\pdfstrcmp{#1}{dec2}=0 0 \else 0\fi\fi\fi\fi\fi \else %
 \ifnum\pdfstrcmp{#2}{seed}=0 0 \else %
 \ifnum\pdfstrcmp{#2}{vt}=0 \ifnum\pdfstrcmp{#1}{dec}=0 1 \else\ifnum\pdfstrcmp{#1}{sci}=0 1 \times 10^{0} \else\ifnum\pdfstrcmp{#1}{dec0}=0 1 \else\ifnum\pdfstrcmp{#1}{dec1}=0 1 \else\ifnum\pdfstrcmp{#1}{dec2}=0 1 \else 1\fi\fi\fi\fi\fi \else %
 \ifnum\pdfstrcmp{#2}{eps_km}=0 \ifnum\pdfstrcmp{#1}{dec}=0 0 \else\ifnum\pdfstrcmp{#1}{sci}=0 0 \else\ifnum\pdfstrcmp{#1}{dec0}=0 0 \else\ifnum\pdfstrcmp{#1}{dec1}=0 0 \else\ifnum\pdfstrcmp{#1}{dec2}=0 0 \else 0\fi\fi\fi\fi\fi \else %
 \ifnum\pdfstrcmp{#2}{dt}=0 \ifnum\pdfstrcmp{#1}{dec}=0 0.001 \else\ifnum\pdfstrcmp{#1}{sci}=0 1 \times 10^{-3} \else\ifnum\pdfstrcmp{#1}{dec0}=0 0 \else\ifnum\pdfstrcmp{#1}{dec1}=0 0 \else\ifnum\pdfstrcmp{#1}{dec2}=0 0 \else 0.001\fi\fi\fi\fi\fi \else %
 \ifnum\pdfstrcmp{#2}{vhist_l}=0 \ifnum\pdfstrcmp{#1}{dec}=0 -0.1 \else\ifnum\pdfstrcmp{#1}{sci}=0 -1 \times 10^{-1} \else\ifnum\pdfstrcmp{#1}{dec0}=0 -0 \else\ifnum\pdfstrcmp{#1}{dec1}=0 -0.1 \else\ifnum\pdfstrcmp{#1}{dec2}=0 -0.1 \else -0.1\fi\fi\fi\fi\fi \else %
 \ifnum\pdfstrcmp{#2}{integ_infty}=0 \ifnum\pdfstrcmp{#1}{dec}=0 15 \else\ifnum\pdfstrcmp{#1}{sci}=0 1.5 \times 10^{1} \else\ifnum\pdfstrcmp{#1}{dec0}=0 15 \else\ifnum\pdfstrcmp{#1}{dec1}=0 15 \else\ifnum\pdfstrcmp{#1}{dec2}=0 15 \else 15\fi\fi\fi\fi\fi \else %
 \ifnum\pdfstrcmp{#2}{integ_epsabs}=0 \ifnum\pdfstrcmp{#1}{dec}=0 0.0001 \else\ifnum\pdfstrcmp{#1}{sci}=0 1 \times 10^{-4} \else\ifnum\pdfstrcmp{#1}{dec0}=0 0 \else\ifnum\pdfstrcmp{#1}{dec1}=0 0 \else\ifnum\pdfstrcmp{#1}{dec2}=0 0 \else 1 \times 10^{-4}\fi\fi\fi\fi\fi \else %
 \ifnum\pdfstrcmp{#2}{_dt_bin}=0 \ifnum\pdfstrcmp{#1}{dec}=0 0.02 \else\ifnum\pdfstrcmp{#1}{sci}=0 2 \times 10^{-2} \else\ifnum\pdfstrcmp{#1}{dec0}=0 0 \else\ifnum\pdfstrcmp{#1}{dec1}=0 0 \else\ifnum\pdfstrcmp{#1}{dec2}=0 0.02 \else 0.02\fi\fi\fi\fi\fi \else %
 \ifnum\pdfstrcmp{#2}{f_c}=0 \ifnum\pdfstrcmp{#1}{dec}=0 1 \else\ifnum\pdfstrcmp{#1}{sci}=0 1 \times 10^{0} \else\ifnum\pdfstrcmp{#1}{dec0}=0 1 \else\ifnum\pdfstrcmp{#1}{dec1}=0 1 \else\ifnum\pdfstrcmp{#1}{dec2}=0 1 \else 1\fi\fi\fi\fi\fi \else %
 \ifnum\pdfstrcmp{#2}{tr}=0 \ifnum\pdfstrcmp{#1}{dec}=0 0.1 \else\ifnum\pdfstrcmp{#1}{sci}=0 1 \times 10^{-1} \else\ifnum\pdfstrcmp{#1}{dec0}=0 0 \else\ifnum\pdfstrcmp{#1}{dec1}=0 0.1 \else\ifnum\pdfstrcmp{#1}{dec2}=0 0.1 \else 0.1\fi\fi\fi\fi\fi \else %
 \ifnum\pdfstrcmp{#2}{km}=0 \ifnum\pdfstrcmp{#1}{dec}=0 2 \else\ifnum\pdfstrcmp{#1}{sci}=0 2 \times 10^{0} \else\ifnum\pdfstrcmp{#1}{dec0}=0 2 \else\ifnum\pdfstrcmp{#1}{dec1}=0 2 \else\ifnum\pdfstrcmp{#1}{dec2}=0 2 \else 2\fi\fi\fi\fi\fi \else %
 \ifnum\pdfstrcmp{#2}{mu}=0 \ifnum\pdfstrcmp{#1}{dec}=0 0.8 \else\ifnum\pdfstrcmp{#1}{sci}=0 8 \times 10^{-1} \else\ifnum\pdfstrcmp{#1}{dec0}=0 1 \else\ifnum\pdfstrcmp{#1}{dec1}=0 0.8 \else\ifnum\pdfstrcmp{#1}{dec2}=0 0.8 \else 0.8\fi\fi\fi\fi\fi \else %
 \ifnum\pdfstrcmp{#2}{s}=0 \ifnum\pdfstrcmp{#1}{dec}=0 2.4 \else\ifnum\pdfstrcmp{#1}{sci}=0 2.4 \times 10^{0} \else\ifnum\pdfstrcmp{#1}{dec0}=0 2 \else\ifnum\pdfstrcmp{#1}{dec1}=0 2.4 \else\ifnum\pdfstrcmp{#1}{dec2}=0 2.4 \else 2.4\fi\fi\fi\fi\fi \else %
 \ifnum\pdfstrcmp{#2}{kp}=0 \ifnum\pdfstrcmp{#1}{dec}=0 1 \else\ifnum\pdfstrcmp{#1}{sci}=0 1 \times 10^{0} \else\ifnum\pdfstrcmp{#1}{dec0}=0 1 \else\ifnum\pdfstrcmp{#1}{dec1}=0 1 \else\ifnum\pdfstrcmp{#1}{dec2}=0 1 \else 1\fi\fi\fi\fi\fi \else %
 \ifnum\pdfstrcmp{#2}{model}=0 lif \else %
 \ifnum\pdfstrcmp{#2}{f_sig}=0 \ifnum\pdfstrcmp{#1}{dec}=0 -1 \else\ifnum\pdfstrcmp{#1}{sci}=0 -1 \times 10^{0} \else\ifnum\pdfstrcmp{#1}{dec0}=0 -1 \else\ifnum\pdfstrcmp{#1}{dec1}=0 -1 \else\ifnum\pdfstrcmp{#1}{dec2}=0 -1 \else -1\fi\fi\fi\fi\fi \else %
 \ifnum\pdfstrcmp{#2}{vhist_N}=0 100 \else %
 \PackageError{paramvalue}{unknown param name: #2}{} 
 \fi\fi\fi\fi\fi\fi\fi\fi\fi\fi\fi\fi\fi\fi\fi\fi\fi\fi\fi\fi\fi\fi\fi\fi}
 
 \includegraphics[scale=1]{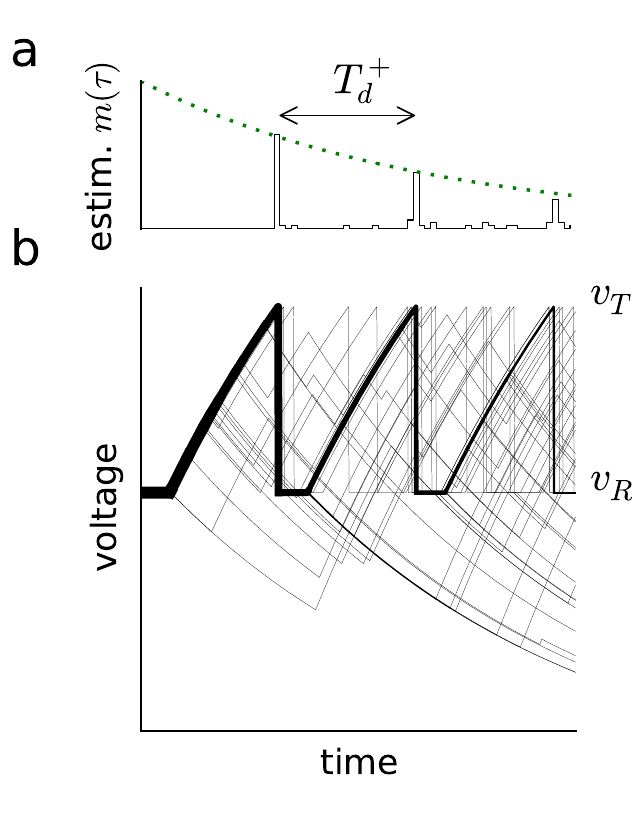}\includegraphics[scale=1]{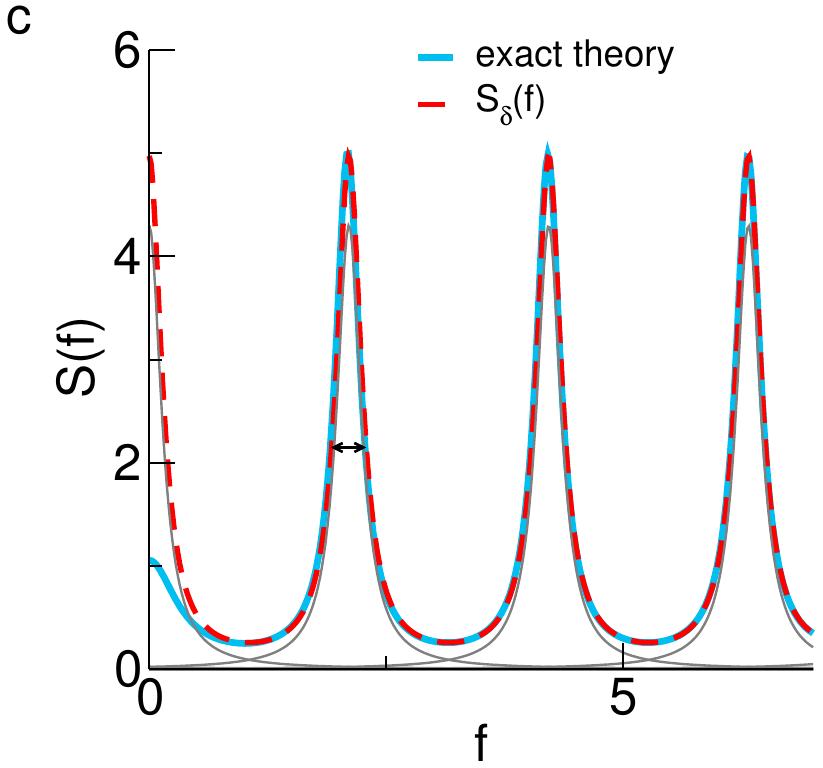}
 \caption{ \label{fig:discspec} {\bf $\delta$ peaks in the spike-triggered rate cause the periodic structure of the power spectrum} 
 {\bf (a)} Estimate of the spike-triggered rate $m(\tau)$ from an ensemble of $\paramvalue{_N_trajs}$ voltage trajectories with different noise realizations that all start in a refractory state at $\tau=0$. {\bf (b)} Corresponding voltage traces. The line thickness is proportional to the number of overlapping trajectories. The estimate for $m(\tau)$ is obtained by binning threshold crossings (bin width $\Delta t = \paramvalue{_dt_bin}$). The green dotted line marks the weight of the $\delta$ peak contributions to $m(\tau)$, given by $\exp[-\hat{k}_+\tau]$ (\e{khat}). {\bf (c)} $S_{\delta}(f)$ , the part of the power spectrum that is due to $\delta$ peaks in the spike-triggered rate (red dashed line, \e{s_delta}), compared to the full expression (blue solid line, \e{powspec}). Also shown are the Lorentzians that are superposed to obtain $S_{\delta}(f)$ (gray lines). The length of the arrow, which marks the full width at half maximum, is $\hat{k}_+/\pi$. Parameters: $k_+ = \paramvalue{kp}, k_- = \paramvalue{km}, \mu = \paramvalue{mu}, \tr=\paramvalue{tr}, \s=\paramvalue{s}, v_R=\paramvalue{vr}, v_T=\paramvalue{vt}$, $\eps=\paramvalue{eps_mu}$.}
\end{figure*}

To understand in a more quantitative way how the ongoing oscillation in the power spectrum arises, consider the structure of the spike-triggered rate.
There is a certain probability that after a neuron has fired, the noise does not switch but remains in the plus state long enough for the neuron to cross the threshold again.
Due to the absence of further stochasticity within the two states, this means that a non-vanishing fraction of trajectories that have been reset at $\tau=0$ crosses the threshold again \emph{exactly} at $\tau=T_d^+$ (see \fig{discspec}b).
In the spike triggered rate, these trajectories become manifest as a $\delta$ peak at the deterministic time from reset to threshold, $T_d^+$ (\fig{discspec}a).
Of course, albeit smaller, there is also a non-vanishing probability that the noise stays in the plus state until these trajectories hit the threshold a second time at $\tau=2 T_d^+$, and so on.
The spike-triggered rate can thus be split into a part containing $\delta$ functions and a continuous part,
\begin{equation}
 m(\tau) = m_{\delta}(\tau) + m_{\rm cont}(\tau).
\end{equation}

The fraction of trajectories contributing to the first $\delta$ peak in $m(\tau)$ is determined as follows:
After the reference spike, all trajectories are clamped at $v_R$ during the refractory period $\tr$. 
During this time, the noise may switch, provided it switches often enough to end up in the plus state after $\tr$.
The fraction of trajectories for which the noise then remains in the plus state between $\tr$ and $T_d^+$ is given by $\exp[-k_+(T_d^+-\tr)]$.
One thus has
\begin{equation}
\begin{split}
 m_{\delta}(\tau) = & \suml_{n=1}^{\infty} \left( P_{+|+}(\tr) \cdot e^{- k_+ (T_d^+-\tr)} \right)^{n} \delta(\tau - n T_d^+),
 \end{split}
\end{equation}
or
\begin{equation}
\begin{split}
 m_{\delta}(\tau) &= \theta(\tau) e^{- \hat{k}_+\tau} \left[\sha_{T_d^+}(\tau) - \delta(\tau) \right],
\end{split}
\end{equation}
where $\sha_{T_d^+}(\tau) := \sum_{n=-\infty}^{\infty} \delta(\tau - n T_d^+)$ is a Dirac comb and $\hat{k}_+$ is given by \e{khat}.
As the Fourier transform of a Dirac comb is again a Dirac comb and multiplication in the time domain turns into convolution in the Fourier domain, one obtains for the part of the power spectrum that is due to the $m_{\delta}(\tau)$,
\begin{equation}
 \begin{split}
 \label{eq:s_delta}
 S_{\delta}(f) &= r_0 \left(1 + 2 \Re \left[\frac{1}{\hat{k}_+ - 2\pi i f} \ast \left(\frac{1}{T_d^+} \sha_{1/T_d^+}(f) - 1 \right)\right]\right) \\
 &= \frac{r_0}{T_d^+} \suml_{n=-\infty}^{\infty} \frac{2 \hat{k}_+}{\hat{k}_+^2 + \left(2 \pi \left[f - \frac{n}{T_d^+} \right]\right)^2},
 \end{split} 
\end{equation}
where $\ast$ denotes convolution.
This is a superposition of Lorentzians positioned at multiples of $1/T_d^+$ in frequency space, the full-width-at-half-maximum of which is given by $\hat{k}_+/\pi$.
Although, at first glance, \e{s_delta} looks quite different from high-frequency limit \e{powspec_highfreq} (unsurprisingly, given how they were derived), the two expressions are equivalent (see \app{highfreq_equiv}).

In \fig{discspec}c, we compare $S_{\delta}(f)$ to the full power spectrum \e{powspec}. 
At higher frequencies, the agreement is excellent, meaning that there the spectrum is dominated by the $\delta$ contributions.

Note that the results regarding $S_\delta(f)$ were derived without using assumptions specific to the LIF neuron model. 
One can thus conclude that the spike train spectra of all dichotomous-noise-driven neurons exhibit a periodic structure given by \e{s_delta} (the specific model only enters via $r_0$ and $T_d^+$), as long as the dichotomous noise is the only source of stochasticity and the parameter regime is such that firing occurs only in the plus state and is regular. 

\section{Susceptibility}
\label{sec:suscep}

How does a weak time-dependent stimulus modulate the instantaneous firing rate of a neuron?
This question can be approached using linear response theory, i.e. assuming that the stimulus is sufficiently weak that its effect on the rate can be described by convolution with a kernel $K(\tau)$,
\begin{equation}
 \label{eq:linear_response_ansatz}
 r(t) \approx r_0 + \varepsilon \intl_{-\infty}^{\infty} d\tau\; K(\tau) s(t-\tau).
\end{equation}
In the following, we calculate the susceptibility $\chi(f)$, the Fourier transform of the linear response kernel $K(\tau)$, for a current stimulus (for signals that enter in a different way, e.g. as a modulation in the switching rate, $\chi(f)$ can be calculated in a similar manner, see \citep{Dro15}).
In linear response, it is sufficient to consider the response to a periodic stimulus, as is evident by plugging $s(t) = \exp(-2\pi i f t)$ into \e{linear_response_ansatz}, which yields
\begin{equation}
 \label{eq:r_ansatz} 
 r(t) = r_0 + \chi(f) \varepsilon e^{-2\pi i f t}.
\end{equation}
The ansatz \e{r_ansatz}, together with the assumption of a cyclostationary solution,
\begin{equation}
 P_{\pm}(v,t) = P_{\pm,0}(v) + \varepsilon e^{-2\pi i f t} P_{\pm,1}(v,f) + \mathcal{O}(\varepsilon^2) \label{eq:p_ansatz},
\end{equation} 
can be plugged into the master equation, \es{modulated_mastereq_pp}{modulated_mastereq_pm}. Keeping only the terms linear in $\eps$, this yields a tractable problem. 
As shown in \app{susdetails}, one obtains equations of the same form as for the power spectrum; in particular, $\chi(f)$ is extracted from \e{jzerosol}, using only different inhomogeneities,
\begin{align}
 \begin{split}
 \tDelta_+(z) &= \frac{\chi(f)}{2\s} \left[e^{2\pi i f\tr}  P_{+|+}(\tr) \delta(x - v_R) - \delta(x - v_T ) \right]\\
 &\quad - \frac{1}{4\s^2} \frac{r_0}{2\pi i f - 1} \left[ P_{+|+}(\tr) \delta'(x - v_R) - \delta'(x - v_T) \right], 
 \end{split} \\
 \tDelta_-(z) &= \frac{\chi(f)}{2\s}  e^{2\pi i f\tr}  P_{-|+}(\tr) \delta(x - v_R) - \frac{1}{4\s^2} \frac{r_0}{2\pi i f - 1} P_{-|+}(\tr) \delta'(x - v_R).
\end{align}
We obtain
\begin{align}
 \label{eq:suscep}
 \chi(f) &= -\frac{r_0}{2\s} \frac{1}{2\pi i f-1} \frac{ \mathcal{F}'(z_T,f) -  P_{+|+}(\tr) \mathcal{F}'(z_R,f) - \frac{k_- P_{-|+}(\tr)}{k_- - 2\pi i f}  \mathcal{G}'(z_R,f)}{\mathcal{F}(z_T,f) - e^{2\pi i f\tr} \left[P_{+|+}(\tr) \mathcal{F}(z_R,f) + \frac{k_- P_{-|+}(\tr)}{k_- - 2\pi i f}  \mathcal{G}(z_R,f) \right]},
\end{align}
where $\mathcal{F}'(\bar{z},f) = \partial_z \mathcal{F}(z,f)|_{\bar{z}}$ and $\mathcal{G}'(z,f)= \partial_z \mathcal{G}(z,f)|_{\bar{z}}$ (see \app{susdetails}).
This is the second central result of this work.

For vanishing refractory period, \e{suscep} can again be written more compactly,
\begin{align}
 \chi(f) &= -\frac{r_0}{2\s} \frac{1}{2\pi i f-1} \frac{ \mathcal{F}'(z_T,f) -  \mathcal{F}'(z_R,f)}{ \mathcal{F}(z_T,f) -  \mathcal{F}(z_R,f)}.
\end{align}
This expression is of the same form as that for the susceptibility of an LIF driven by white noise as given in ref. \citep{BruCha01}.

\begin{figure}
 \newcommand{\paramvalue}[2][]{\protect %
 \ifnum\pdfstrcmp{#2}{integ_epsabs}=0 \ifnum\pdfstrcmp{#1}{dec}=0 0.0001 \else\ifnum\pdfstrcmp{#1}{sci}=0 1 \times 10^{-4} \else\ifnum\pdfstrcmp{#1}{dec0}=0 0 \else\ifnum\pdfstrcmp{#1}{dec1}=0 0 \else\ifnum\pdfstrcmp{#1}{dec2}=0 0 \else 1 \times 10^{-4}\fi\fi\fi\fi\fi \else %
 \ifnum\pdfstrcmp{#2}{vhist_r}=0 \ifnum\pdfstrcmp{#1}{dec}=0 1.1 \else\ifnum\pdfstrcmp{#1}{sci}=0 1.1 \times 10^{0} \else\ifnum\pdfstrcmp{#1}{dec0}=0 1 \else\ifnum\pdfstrcmp{#1}{dec1}=0 1.1 \else\ifnum\pdfstrcmp{#1}{dec2}=0 1.1 \else 1.1\fi\fi\fi\fi\fi \else %
 \ifnum\pdfstrcmp{#2}{integ_infty}=0 \ifnum\pdfstrcmp{#1}{dec}=0 15 \else\ifnum\pdfstrcmp{#1}{sci}=0 1.5 \times 10^{1} \else\ifnum\pdfstrcmp{#1}{dec0}=0 15 \else\ifnum\pdfstrcmp{#1}{dec1}=0 15 \else\ifnum\pdfstrcmp{#1}{dec2}=0 15 \else 15\fi\fi\fi\fi\fi \else %
 \ifnum\pdfstrcmp{#2}{N_trials}=0 10000 \else %
 \ifnum\pdfstrcmp{#2}{df}=0 \ifnum\pdfstrcmp{#1}{dec}=0 0.01 \else\ifnum\pdfstrcmp{#1}{sci}=0 1 \times 10^{-2} \else\ifnum\pdfstrcmp{#1}{dec0}=0 0 \else\ifnum\pdfstrcmp{#1}{dec1}=0 0 \else\ifnum\pdfstrcmp{#1}{dec2}=0 0.01 \else 0.01\fi\fi\fi\fi\fi \else %
 \ifnum\pdfstrcmp{#2}{vr}=0 \ifnum\pdfstrcmp{#1}{dec}=0 0 \else\ifnum\pdfstrcmp{#1}{sci}=0 0 \else\ifnum\pdfstrcmp{#1}{dec0}=0 0 \else\ifnum\pdfstrcmp{#1}{dec1}=0 0 \else\ifnum\pdfstrcmp{#1}{dec2}=0 0 \else 0\fi\fi\fi\fi\fi \else %
 \ifnum\pdfstrcmp{#2}{seed}=0 eval_after_unroll(int(f_sig*11234)) \else %
 \ifnum\pdfstrcmp{#2}{vt}=0 \ifnum\pdfstrcmp{#1}{dec}=0 1 \else\ifnum\pdfstrcmp{#1}{sci}=0 1 \times 10^{0} \else\ifnum\pdfstrcmp{#1}{dec0}=0 1 \else\ifnum\pdfstrcmp{#1}{dec1}=0 1 \else\ifnum\pdfstrcmp{#1}{dec2}=0 1 \else 1\fi\fi\fi\fi\fi \else %
 \ifnum\pdfstrcmp{#2}{eps_km}=0 \ifnum\pdfstrcmp{#1}{dec}=0 0 \else\ifnum\pdfstrcmp{#1}{sci}=0 0 \else\ifnum\pdfstrcmp{#1}{dec0}=0 0 \else\ifnum\pdfstrcmp{#1}{dec1}=0 0 \else\ifnum\pdfstrcmp{#1}{dec2}=0 0 \else 0\fi\fi\fi\fi\fi \else %
 \ifnum\pdfstrcmp{#2}{dt}=0 \ifnum\pdfstrcmp{#1}{dec}=0 0.0001 \else\ifnum\pdfstrcmp{#1}{sci}=0 1 \times 10^{-4} \else\ifnum\pdfstrcmp{#1}{dec0}=0 0 \else\ifnum\pdfstrcmp{#1}{dec1}=0 0 \else\ifnum\pdfstrcmp{#1}{dec2}=0 0 \else 1 \times 10^{-4}\fi\fi\fi\fi\fi \else %
 \ifnum\pdfstrcmp{#2}{vhist_l}=0 \ifnum\pdfstrcmp{#1}{dec}=0 -0.1 \else\ifnum\pdfstrcmp{#1}{sci}=0 -1 \times 10^{-1} \else\ifnum\pdfstrcmp{#1}{dec0}=0 -0 \else\ifnum\pdfstrcmp{#1}{dec1}=0 -0.1 \else\ifnum\pdfstrcmp{#1}{dec2}=0 -0.1 \else -0.1\fi\fi\fi\fi\fi \else %
 \ifnum\pdfstrcmp{#2}{eps_mu}=0 \ifnum\pdfstrcmp{#1}{dec}=0 0.2 \else\ifnum\pdfstrcmp{#1}{sci}=0 2 \times 10^{-1} \else\ifnum\pdfstrcmp{#1}{dec0}=0 0 \else\ifnum\pdfstrcmp{#1}{dec1}=0 0.2 \else\ifnum\pdfstrcmp{#1}{dec2}=0 0.2 \else 0.2\fi\fi\fi\fi\fi \else %
 \ifnum\pdfstrcmp{#2}{f_max}=0 \ifnum\pdfstrcmp{#1}{dec}=0 5000 \else\ifnum\pdfstrcmp{#1}{sci}=0 5 \times 10^{3} \else\ifnum\pdfstrcmp{#1}{dec0}=0 5000 \else\ifnum\pdfstrcmp{#1}{dec1}=0 5000 \else\ifnum\pdfstrcmp{#1}{dec2}=0 5000 \else 5000\fi\fi\fi\fi\fi \else %
 \ifnum\pdfstrcmp{#2}{f_c}=0 \ifnum\pdfstrcmp{#1}{dec}=0 1 \else\ifnum\pdfstrcmp{#1}{sci}=0 1 \times 10^{0} \else\ifnum\pdfstrcmp{#1}{dec0}=0 1 \else\ifnum\pdfstrcmp{#1}{dec1}=0 1 \else\ifnum\pdfstrcmp{#1}{dec2}=0 1 \else 1\fi\fi\fi\fi\fi \else %
 \ifnum\pdfstrcmp{#2}{tr}=0 \ifnum\pdfstrcmp{#1}{dec}=0 0 \else\ifnum\pdfstrcmp{#1}{sci}=0 0 \else\ifnum\pdfstrcmp{#1}{dec0}=0 0 \else\ifnum\pdfstrcmp{#1}{dec1}=0 0 \else\ifnum\pdfstrcmp{#1}{dec2}=0 0 \else 0\fi\fi\fi\fi\fi \else %
 \ifnum\pdfstrcmp{#2}{km}=0 \ifnum\pdfstrcmp{#1}{dec}=0 2 \else\ifnum\pdfstrcmp{#1}{sci}=0 2 \times 10^{0} \else\ifnum\pdfstrcmp{#1}{dec0}=0 2 \else\ifnum\pdfstrcmp{#1}{dec1}=0 2 \else\ifnum\pdfstrcmp{#1}{dec2}=0 2 \else 2\fi\fi\fi\fi\fi \else %
 \ifnum\pdfstrcmp{#2}{mu}=0 \ifnum\pdfstrcmp{#1}{dec}=0 0.8 \else\ifnum\pdfstrcmp{#1}{sci}=0 8 \times 10^{-1} \else\ifnum\pdfstrcmp{#1}{dec0}=0 1 \else\ifnum\pdfstrcmp{#1}{dec1}=0 0.8 \else\ifnum\pdfstrcmp{#1}{dec2}=0 0.8 \else 0.8\fi\fi\fi\fi\fi \else %
 \ifnum\pdfstrcmp{#2}{s}=0 \ifnum\pdfstrcmp{#1}{dec}=0 2.4 \else\ifnum\pdfstrcmp{#1}{sci}=0 2.4 \times 10^{0} \else\ifnum\pdfstrcmp{#1}{dec0}=0 2 \else\ifnum\pdfstrcmp{#1}{dec1}=0 2.4 \else\ifnum\pdfstrcmp{#1}{dec2}=0 2.4 \else 2.4\fi\fi\fi\fi\fi \else %
 \ifnum\pdfstrcmp{#2}{kp}=0 \ifnum\pdfstrcmp{#1}{dec}=0 1 \else\ifnum\pdfstrcmp{#1}{sci}=0 1 \times 10^{0} \else\ifnum\pdfstrcmp{#1}{dec0}=0 1 \else\ifnum\pdfstrcmp{#1}{dec1}=0 1 \else\ifnum\pdfstrcmp{#1}{dec2}=0 1 \else 1\fi\fi\fi\fi\fi \else %
 \ifnum\pdfstrcmp{#2}{model}=0 lif \else %
 \ifnum\pdfstrcmp{#2}{f_sig}=0 [  1.00000000e-02   2.11919192e-01   4.13838384e-01   6.15757576e-01
    8.17676768e-01   1.01959596e+00   1.22151515e+00   1.42343434e+00
    1.62535354e+00   1.82727273e+00   2.02919192e+00   2.23111111e+00
    2.43303030e+00   2.63494949e+00   2.83686869e+00   3.03878788e+00
    3.24070707e+00   3.44262626e+00   3.64454545e+00   3.84646465e+00
    4.04838384e+00   4.25030303e+00   4.45222222e+00   4.65414141e+00
    4.85606061e+00   5.05797980e+00   5.25989899e+00   5.46181818e+00
    5.66373737e+00   5.86565657e+00   6.06757576e+00   6.26949495e+00
    6.47141414e+00   6.67333333e+00   6.87525253e+00   7.07717172e+00
    7.27909091e+00   7.48101010e+00   7.68292929e+00   7.88484848e+00
    8.08676768e+00   8.28868687e+00   8.49060606e+00   8.69252525e+00
    8.89444444e+00   9.09636364e+00   9.29828283e+00   9.50020202e+00
    9.70212121e+00   9.90404040e+00   1.01059596e+01   1.03078788e+01
    1.05097980e+01   1.07117172e+01   1.09136364e+01   1.11155556e+01
    1.13174747e+01   1.15193939e+01   1.17213131e+01   1.19232323e+01
    1.21251515e+01   1.23270707e+01   1.25289899e+01   1.27309091e+01
    1.29328283e+01   1.31347475e+01   1.33366667e+01   1.35385859e+01
    1.37405051e+01   1.39424242e+01   1.41443434e+01   1.43462626e+01
    1.45481818e+01   1.47501010e+01   1.49520202e+01   1.51539394e+01
    1.53558586e+01   1.55577778e+01   1.57596970e+01   1.59616162e+01
    1.61635354e+01   1.63654545e+01   1.65673737e+01   1.67692929e+01
    1.69712121e+01   1.71731313e+01   1.73750505e+01   1.75769697e+01
    1.77788889e+01   1.79808081e+01   1.81827273e+01   1.83846465e+01
    1.85865657e+01   1.87884848e+01   1.89904040e+01   1.91923232e+01
    1.93942424e+01   1.95961616e+01   1.97980808e+01   2.00000000e+01] \else %
 \ifnum\pdfstrcmp{#2}{vhist_N}=0 100 \else %
 \PackageError{paramvalue}{unknown param name: #2}{} 
 \fi\fi\fi\fi\fi\fi\fi\fi\fi\fi\fi\fi\fi\fi\fi\fi\fi\fi\fi\fi\fi\fi}
 
 \includegraphics{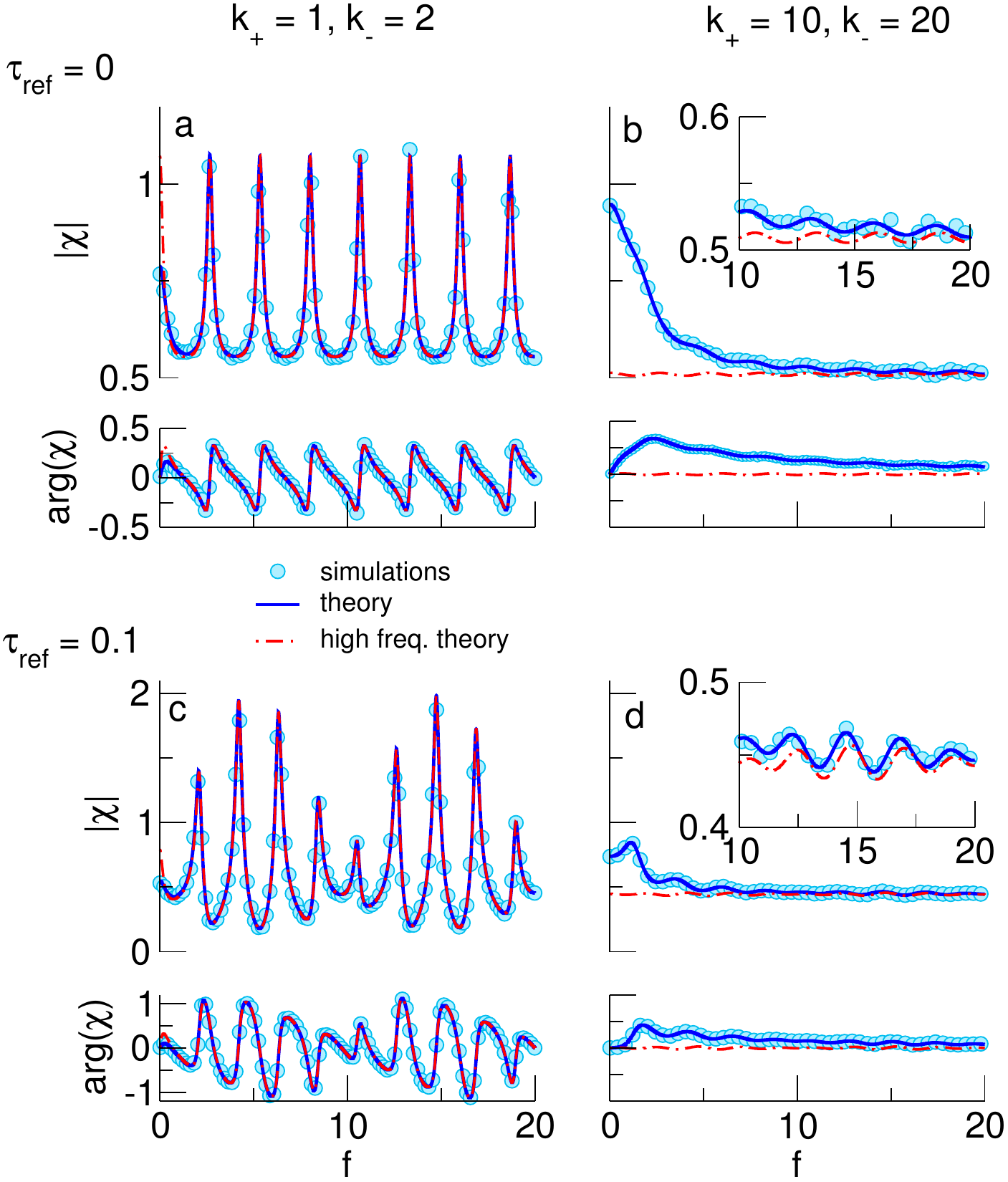}
 \caption{\label{fig:suscep} {\bf Susceptibility of an LIF neuron driven by dichotomous noise.} Shown are the absolute value and the complex phase for two different switching rate combinations, once without refractory period and once for $\tr=0.1$. The insets in (b) and (d) show a zoomed version of the same curves at higher frequencies. Each symbol represents the result of the simulation of an LIF stimulated with a sinusoidal signal at that frequency. Remaining parameters: $\mu = \paramvalue{mu}, \s=\paramvalue{s}, v_R=\paramvalue{vr}, v_T=\paramvalue{vt}$, $\eps=\paramvalue{eps_mu}$.}
\end{figure}

{\newcommand{\paramvalue}[2][]{\protect %
\ifnum\pdfstrcmp{#2}{integ_epsabs}=0 \ifnum\pdfstrcmp{#1}{dec}=0 0.0001 \else\ifnum\pdfstrcmp{#1}{sci}=0 1 \times 10^{-4} \else\ifnum\pdfstrcmp{#1}{dec0}=0 0 \else\ifnum\pdfstrcmp{#1}{dec1}=0 0 \else\ifnum\pdfstrcmp{#1}{dec2}=0 0 \else 1 \times 10^{-4}\fi\fi\fi\fi\fi \else %
\ifnum\pdfstrcmp{#2}{vhist_r}=0 \ifnum\pdfstrcmp{#1}{dec}=0 1.1 \else\ifnum\pdfstrcmp{#1}{sci}=0 1.1 \times 10^{0} \else\ifnum\pdfstrcmp{#1}{dec0}=0 1 \else\ifnum\pdfstrcmp{#1}{dec1}=0 1.1 \else\ifnum\pdfstrcmp{#1}{dec2}=0 1.1 \else 1.1\fi\fi\fi\fi\fi \else %
\ifnum\pdfstrcmp{#2}{integ_infty}=0 \ifnum\pdfstrcmp{#1}{dec}=0 15 \else\ifnum\pdfstrcmp{#1}{sci}=0 1.5 \times 10^{1} \else\ifnum\pdfstrcmp{#1}{dec0}=0 15 \else\ifnum\pdfstrcmp{#1}{dec1}=0 15 \else\ifnum\pdfstrcmp{#1}{dec2}=0 15 \else 15\fi\fi\fi\fi\fi \else %
\ifnum\pdfstrcmp{#2}{N_trials}=0 10000 \else %
\ifnum\pdfstrcmp{#2}{df}=0 \ifnum\pdfstrcmp{#1}{dec}=0 0.01 \else\ifnum\pdfstrcmp{#1}{sci}=0 1 \times 10^{-2} \else\ifnum\pdfstrcmp{#1}{dec0}=0 0 \else\ifnum\pdfstrcmp{#1}{dec1}=0 0 \else\ifnum\pdfstrcmp{#1}{dec2}=0 0.01 \else 0.01\fi\fi\fi\fi\fi \else %
\ifnum\pdfstrcmp{#2}{vr}=0 \ifnum\pdfstrcmp{#1}{dec}=0 0 \else\ifnum\pdfstrcmp{#1}{sci}=0 0 \else\ifnum\pdfstrcmp{#1}{dec0}=0 0 \else\ifnum\pdfstrcmp{#1}{dec1}=0 0 \else\ifnum\pdfstrcmp{#1}{dec2}=0 0 \else 0\fi\fi\fi\fi\fi \else %
\ifnum\pdfstrcmp{#2}{seed}=0 eval_after_unroll(int(f_sig*11234)) \else %
\ifnum\pdfstrcmp{#2}{vt}=0 \ifnum\pdfstrcmp{#1}{dec}=0 1 \else\ifnum\pdfstrcmp{#1}{sci}=0 1 \times 10^{0} \else\ifnum\pdfstrcmp{#1}{dec0}=0 1 \else\ifnum\pdfstrcmp{#1}{dec1}=0 1 \else\ifnum\pdfstrcmp{#1}{dec2}=0 1 \else 1\fi\fi\fi\fi\fi \else %
\ifnum\pdfstrcmp{#2}{eps_km}=0 \ifnum\pdfstrcmp{#1}{dec}=0 0 \else\ifnum\pdfstrcmp{#1}{sci}=0 0 \else\ifnum\pdfstrcmp{#1}{dec0}=0 0 \else\ifnum\pdfstrcmp{#1}{dec1}=0 0 \else\ifnum\pdfstrcmp{#1}{dec2}=0 0 \else 0\fi\fi\fi\fi\fi \else %
\ifnum\pdfstrcmp{#2}{dt}=0 \ifnum\pdfstrcmp{#1}{dec}=0 0.0001 \else\ifnum\pdfstrcmp{#1}{sci}=0 1 \times 10^{-4} \else\ifnum\pdfstrcmp{#1}{dec0}=0 0 \else\ifnum\pdfstrcmp{#1}{dec1}=0 0 \else\ifnum\pdfstrcmp{#1}{dec2}=0 0 \else 1 \times 10^{-4}\fi\fi\fi\fi\fi \else %
\ifnum\pdfstrcmp{#2}{vhist_l}=0 \ifnum\pdfstrcmp{#1}{dec}=0 -0.1 \else\ifnum\pdfstrcmp{#1}{sci}=0 -1 \times 10^{-1} \else\ifnum\pdfstrcmp{#1}{dec0}=0 -0 \else\ifnum\pdfstrcmp{#1}{dec1}=0 -0.1 \else\ifnum\pdfstrcmp{#1}{dec2}=0 -0.1 \else -0.1\fi\fi\fi\fi\fi \else %
\ifnum\pdfstrcmp{#2}{eps_mu}=0 \ifnum\pdfstrcmp{#1}{dec}=0 0.2 \else\ifnum\pdfstrcmp{#1}{sci}=0 2 \times 10^{-1} \else\ifnum\pdfstrcmp{#1}{dec0}=0 0 \else\ifnum\pdfstrcmp{#1}{dec1}=0 0.2 \else\ifnum\pdfstrcmp{#1}{dec2}=0 0.2 \else 0.2\fi\fi\fi\fi\fi \else %
\ifnum\pdfstrcmp{#2}{f_max}=0 \ifnum\pdfstrcmp{#1}{dec}=0 5000 \else\ifnum\pdfstrcmp{#1}{sci}=0 5 \times 10^{3} \else\ifnum\pdfstrcmp{#1}{dec0}=0 5000 \else\ifnum\pdfstrcmp{#1}{dec1}=0 5000 \else\ifnum\pdfstrcmp{#1}{dec2}=0 5000 \else 5000\fi\fi\fi\fi\fi \else %
\ifnum\pdfstrcmp{#2}{f_c}=0 \ifnum\pdfstrcmp{#1}{dec}=0 1 \else\ifnum\pdfstrcmp{#1}{sci}=0 1 \times 10^{0} \else\ifnum\pdfstrcmp{#1}{dec0}=0 1 \else\ifnum\pdfstrcmp{#1}{dec1}=0 1 \else\ifnum\pdfstrcmp{#1}{dec2}=0 1 \else 1\fi\fi\fi\fi\fi \else %
\ifnum\pdfstrcmp{#2}{tr}=0 \ifnum\pdfstrcmp{#1}{dec}=0 0 \else\ifnum\pdfstrcmp{#1}{sci}=0 0 \else\ifnum\pdfstrcmp{#1}{dec0}=0 0 \else\ifnum\pdfstrcmp{#1}{dec1}=0 0 \else\ifnum\pdfstrcmp{#1}{dec2}=0 0 \else 0\fi\fi\fi\fi\fi \else %
\ifnum\pdfstrcmp{#2}{km}=0 \ifnum\pdfstrcmp{#1}{dec}=0 2 \else\ifnum\pdfstrcmp{#1}{sci}=0 2 \times 10^{0} \else\ifnum\pdfstrcmp{#1}{dec0}=0 2 \else\ifnum\pdfstrcmp{#1}{dec1}=0 2 \else\ifnum\pdfstrcmp{#1}{dec2}=0 2 \else 2\fi\fi\fi\fi\fi \else %
\ifnum\pdfstrcmp{#2}{mu}=0 \ifnum\pdfstrcmp{#1}{dec}=0 0.8 \else\ifnum\pdfstrcmp{#1}{sci}=0 8 \times 10^{-1} \else\ifnum\pdfstrcmp{#1}{dec0}=0 1 \else\ifnum\pdfstrcmp{#1}{dec1}=0 0.8 \else\ifnum\pdfstrcmp{#1}{dec2}=0 0.8 \else 0.8\fi\fi\fi\fi\fi \else %
\ifnum\pdfstrcmp{#2}{s}=0 \ifnum\pdfstrcmp{#1}{dec}=0 2.4 \else\ifnum\pdfstrcmp{#1}{sci}=0 2.4 \times 10^{0} \else\ifnum\pdfstrcmp{#1}{dec0}=0 2 \else\ifnum\pdfstrcmp{#1}{dec1}=0 2.4 \else\ifnum\pdfstrcmp{#1}{dec2}=0 2.4 \else 2.4\fi\fi\fi\fi\fi \else %
\ifnum\pdfstrcmp{#2}{kp}=0 \ifnum\pdfstrcmp{#1}{dec}=0 1 \else\ifnum\pdfstrcmp{#1}{sci}=0 1 \times 10^{0} \else\ifnum\pdfstrcmp{#1}{dec0}=0 1 \else\ifnum\pdfstrcmp{#1}{dec1}=0 1 \else\ifnum\pdfstrcmp{#1}{dec2}=0 1 \else 1\fi\fi\fi\fi\fi \else %
\ifnum\pdfstrcmp{#2}{model}=0 lif \else %
\ifnum\pdfstrcmp{#2}{f_sig}=0 [  1.00000000e-02   2.11919192e-01   4.13838384e-01   6.15757576e-01
   8.17676768e-01   1.01959596e+00   1.22151515e+00   1.42343434e+00
   1.62535354e+00   1.82727273e+00   2.02919192e+00   2.23111111e+00
   2.43303030e+00   2.63494949e+00   2.83686869e+00   3.03878788e+00
   3.24070707e+00   3.44262626e+00   3.64454545e+00   3.84646465e+00
   4.04838384e+00   4.25030303e+00   4.45222222e+00   4.65414141e+00
   4.85606061e+00   5.05797980e+00   5.25989899e+00   5.46181818e+00
   5.66373737e+00   5.86565657e+00   6.06757576e+00   6.26949495e+00
   6.47141414e+00   6.67333333e+00   6.87525253e+00   7.07717172e+00
   7.27909091e+00   7.48101010e+00   7.68292929e+00   7.88484848e+00
   8.08676768e+00   8.28868687e+00   8.49060606e+00   8.69252525e+00
   8.89444444e+00   9.09636364e+00   9.29828283e+00   9.50020202e+00
   9.70212121e+00   9.90404040e+00   1.01059596e+01   1.03078788e+01
   1.05097980e+01   1.07117172e+01   1.09136364e+01   1.11155556e+01
   1.13174747e+01   1.15193939e+01   1.17213131e+01   1.19232323e+01
   1.21251515e+01   1.23270707e+01   1.25289899e+01   1.27309091e+01
   1.29328283e+01   1.31347475e+01   1.33366667e+01   1.35385859e+01
   1.37405051e+01   1.39424242e+01   1.41443434e+01   1.43462626e+01
   1.45481818e+01   1.47501010e+01   1.49520202e+01   1.51539394e+01
   1.53558586e+01   1.55577778e+01   1.57596970e+01   1.59616162e+01
   1.61635354e+01   1.63654545e+01   1.65673737e+01   1.67692929e+01
   1.69712121e+01   1.71731313e+01   1.73750505e+01   1.75769697e+01
   1.77788889e+01   1.79808081e+01   1.81827273e+01   1.83846465e+01
   1.85865657e+01   1.87884848e+01   1.89904040e+01   1.91923232e+01
   1.93942424e+01   1.95961616e+01   1.97980808e+01   2.00000000e+01] \else %
\ifnum\pdfstrcmp{#2}{vhist_N}=0 100 \else %
\PackageError{paramvalue}{unknown param name: #2}{} 
\fi\fi\fi\fi\fi\fi\fi\fi\fi\fi\fi\fi\fi\fi\fi\fi\fi\fi\fi\fi\fi\fi}

In \fig{suscep}, we compare absolute value and phase of the exact theory, \e{suscep}, to simulations (each symbol quantifies the response of an LIF neuron stimulated with a sinusoid at the indicated frequency).
The theory matches simulations perfectly within line thickness, indicating that the linear-response assumption is fulfilled for the chosen signal strength $\eps = \paramvalue{eps_mu}$. 
Like the power spectrum, the susceptibility displays an undamped periodic structure that is more prominent for low switching rates (\fig{suscep}a,c).
For $\tr=0$, the period of this peaked structure is again given by the inverse time from reset to threshold in the plus state (\fig{suscep}a).
Interestingly, with a non-vanishing refractory period $\tr>0$, these peaks become modulated by a second oscillation with period $1/\tr$ (\fig{suscep}c).
This is also apparent by looking at the high-frequency limit of \e{suscep},
\begin{equation}
 \label{eq:suscep_highf}
 \begin{split}
  \chi(f \gg 1) = \frac{r_0}{2\s} \frac{1 - P_{+|+}(\tr) e^{-(k_+ + 1)(T_d^+ - \tr)} e^{2 \pi i f(T_d^+-\tr)}}{(1-z_T) \left(1-P_{+|+}(\tr)e^{-k_+(T_d^+ - \tr)}e^{2 \pi i f T_d^+} \right)}
 \end{split},
\end{equation}
which is also shown in \fig{suscep} as the red dashed line.
For a non-vanishing refractory period, the two oscillatory terms in \e{suscep_highf}, $e^{2 \pi i f(T_d^+-\tr)}$ and $e^{2 \pi i f T_d^+}$, differ slightly in their frequencies, leading to a beating at frequency $\tr$. 

The periodic structure, along with the beating, is also apparent for higher switching rates, although less pronounced (insets in \fig{suscep}b,d).
Here, it is particularly noticeable that the susceptibility does not decay to zero (with a non-vanishing phase) in the limit of high frequencies, as it would for integrate-and-fire neurons driven by Gaussian white noise, but instead oscillates weakly around a finite real value.
This means that the neuron can respond to signals of arbitrarily high frequency, a result that has been also attained for LIF neurons driven by a different kind of colored noise (an Ornstein-Uhlenbeck process) \citep{BruCha01, FouBru02}.

}


\section{Spectrum and susceptibility under broadband stimulation}
\label{sec:broadband}

\begin{figure}
 \newcommand{\paramvalue}[2][]{\protect %
 \ifnum\pdfstrcmp{#2}{f_max}=0 \ifnum\pdfstrcmp{#1}{dec}=0 100 \else\ifnum\pdfstrcmp{#1}{sci}=0 1 \times 10^{2} \else\ifnum\pdfstrcmp{#1}{dec0}=0 100 \else\ifnum\pdfstrcmp{#1}{dec1}=0 100 \else\ifnum\pdfstrcmp{#1}{dec2}=0 100 \else 100\fi\fi\fi\fi\fi \else %
 \ifnum\pdfstrcmp{#2}{vhist_r}=0 \ifnum\pdfstrcmp{#1}{dec}=0 1.1 \else\ifnum\pdfstrcmp{#1}{sci}=0 1.1 \times 10^{0} \else\ifnum\pdfstrcmp{#1}{dec0}=0 1 \else\ifnum\pdfstrcmp{#1}{dec1}=0 1.1 \else\ifnum\pdfstrcmp{#1}{dec2}=0 1.1 \else 1.1\fi\fi\fi\fi\fi \else %
 \ifnum\pdfstrcmp{#2}{integ_infty}=0 \ifnum\pdfstrcmp{#1}{dec}=0 15 \else\ifnum\pdfstrcmp{#1}{sci}=0 1.5 \times 10^{1} \else\ifnum\pdfstrcmp{#1}{dec0}=0 15 \else\ifnum\pdfstrcmp{#1}{dec1}=0 15 \else\ifnum\pdfstrcmp{#1}{dec2}=0 15 \else 15\fi\fi\fi\fi\fi \else %
 \ifnum\pdfstrcmp{#2}{N_trials}=0 1000000 \else %
 \ifnum\pdfstrcmp{#2}{df}=0 \ifnum\pdfstrcmp{#1}{dec}=0 0.01 \else\ifnum\pdfstrcmp{#1}{sci}=0 1 \times 10^{-2} \else\ifnum\pdfstrcmp{#1}{dec0}=0 0 \else\ifnum\pdfstrcmp{#1}{dec1}=0 0 \else\ifnum\pdfstrcmp{#1}{dec2}=0 0.01 \else 0.01\fi\fi\fi\fi\fi \else %
 \ifnum\pdfstrcmp{#2}{vr}=0 \ifnum\pdfstrcmp{#1}{dec}=0 0 \else\ifnum\pdfstrcmp{#1}{sci}=0 0 \else\ifnum\pdfstrcmp{#1}{dec0}=0 0 \else\ifnum\pdfstrcmp{#1}{dec1}=0 0 \else\ifnum\pdfstrcmp{#1}{dec2}=0 0 \else 0\fi\fi\fi\fi\fi \else %
 \ifnum\pdfstrcmp{#2}{seed}=0 0 \else %
 \ifnum\pdfstrcmp{#2}{vt}=0 \ifnum\pdfstrcmp{#1}{dec}=0 1 \else\ifnum\pdfstrcmp{#1}{sci}=0 1 \times 10^{0} \else\ifnum\pdfstrcmp{#1}{dec0}=0 1 \else\ifnum\pdfstrcmp{#1}{dec1}=0 1 \else\ifnum\pdfstrcmp{#1}{dec2}=0 1 \else 1\fi\fi\fi\fi\fi \else %
 \ifnum\pdfstrcmp{#2}{eps_km}=0 \ifnum\pdfstrcmp{#1}{dec}=0 0 \else\ifnum\pdfstrcmp{#1}{sci}=0 0 \else\ifnum\pdfstrcmp{#1}{dec0}=0 0 \else\ifnum\pdfstrcmp{#1}{dec1}=0 0 \else\ifnum\pdfstrcmp{#1}{dec2}=0 0 \else 0\fi\fi\fi\fi\fi \else %
 \ifnum\pdfstrcmp{#2}{dt}=0 \ifnum\pdfstrcmp{#1}{dec}=0 0.005 \else\ifnum\pdfstrcmp{#1}{sci}=0 5 \times 10^{-3} \else\ifnum\pdfstrcmp{#1}{dec0}=0 0 \else\ifnum\pdfstrcmp{#1}{dec1}=0 0 \else\ifnum\pdfstrcmp{#1}{dec2}=0 0.01 \else 0.005\fi\fi\fi\fi\fi \else %
 \ifnum\pdfstrcmp{#2}{vhist_l}=0 \ifnum\pdfstrcmp{#1}{dec}=0 -0.1 \else\ifnum\pdfstrcmp{#1}{sci}=0 -1 \times 10^{-1} \else\ifnum\pdfstrcmp{#1}{dec0}=0 -0 \else\ifnum\pdfstrcmp{#1}{dec1}=0 -0.1 \else\ifnum\pdfstrcmp{#1}{dec2}=0 -0.1 \else -0.1\fi\fi\fi\fi\fi \else %
 \ifnum\pdfstrcmp{#2}{eps_mu}=0 \ifnum\pdfstrcmp{#1}{dec}=0 0.7 \else\ifnum\pdfstrcmp{#1}{sci}=0 7 \times 10^{-1} \else\ifnum\pdfstrcmp{#1}{dec0}=0 1 \else\ifnum\pdfstrcmp{#1}{dec1}=0 0.7 \else\ifnum\pdfstrcmp{#1}{dec2}=0 0.7 \else 0.7\fi\fi\fi\fi\fi \else %
 \ifnum\pdfstrcmp{#2}{integ_epsabs}=0 \ifnum\pdfstrcmp{#1}{dec}=0 0.0001 \else\ifnum\pdfstrcmp{#1}{sci}=0 1 \times 10^{-4} \else\ifnum\pdfstrcmp{#1}{dec0}=0 0 \else\ifnum\pdfstrcmp{#1}{dec1}=0 0 \else\ifnum\pdfstrcmp{#1}{dec2}=0 0 \else 1 \times 10^{-4}\fi\fi\fi\fi\fi \else %
 \ifnum\pdfstrcmp{#2}{f_c}=0 \ifnum\pdfstrcmp{#1}{dec}=0 100 \else\ifnum\pdfstrcmp{#1}{sci}=0 1 \times 10^{2} \else\ifnum\pdfstrcmp{#1}{dec0}=0 100 \else\ifnum\pdfstrcmp{#1}{dec1}=0 100 \else\ifnum\pdfstrcmp{#1}{dec2}=0 100 \else 100\fi\fi\fi\fi\fi \else %
 \ifnum\pdfstrcmp{#2}{tr}=0 \ifnum\pdfstrcmp{#1}{dec}=0 0.1 \else\ifnum\pdfstrcmp{#1}{sci}=0 1 \times 10^{-1} \else\ifnum\pdfstrcmp{#1}{dec0}=0 0 \else\ifnum\pdfstrcmp{#1}{dec1}=0 0.1 \else\ifnum\pdfstrcmp{#1}{dec2}=0 0.1 \else 0.1\fi\fi\fi\fi\fi \else %
 \ifnum\pdfstrcmp{#2}{km}=0 \ifnum\pdfstrcmp{#1}{dec}=0 20 \else\ifnum\pdfstrcmp{#1}{sci}=0 2 \times 10^{1} \else\ifnum\pdfstrcmp{#1}{dec0}=0 20 \else\ifnum\pdfstrcmp{#1}{dec1}=0 20 \else\ifnum\pdfstrcmp{#1}{dec2}=0 20 \else 20\fi\fi\fi\fi\fi \else %
 \ifnum\pdfstrcmp{#2}{mu}=0 \ifnum\pdfstrcmp{#1}{dec}=0 0.8 \else\ifnum\pdfstrcmp{#1}{sci}=0 8 \times 10^{-1} \else\ifnum\pdfstrcmp{#1}{dec0}=0 1 \else\ifnum\pdfstrcmp{#1}{dec1}=0 0.8 \else\ifnum\pdfstrcmp{#1}{dec2}=0 0.8 \else 0.8\fi\fi\fi\fi\fi \else %
 \ifnum\pdfstrcmp{#2}{s}=0 \ifnum\pdfstrcmp{#1}{dec}=0 2.4 \else\ifnum\pdfstrcmp{#1}{sci}=0 2.4 \times 10^{0} \else\ifnum\pdfstrcmp{#1}{dec0}=0 2 \else\ifnum\pdfstrcmp{#1}{dec1}=0 2.4 \else\ifnum\pdfstrcmp{#1}{dec2}=0 2.4 \else 2.4\fi\fi\fi\fi\fi \else %
 \ifnum\pdfstrcmp{#2}{kp}=0 \ifnum\pdfstrcmp{#1}{dec}=0 10 \else\ifnum\pdfstrcmp{#1}{sci}=0 1 \times 10^{1} \else\ifnum\pdfstrcmp{#1}{dec0}=0 10 \else\ifnum\pdfstrcmp{#1}{dec1}=0 10 \else\ifnum\pdfstrcmp{#1}{dec2}=0 10 \else 10\fi\fi\fi\fi\fi \else %
 \ifnum\pdfstrcmp{#2}{model}=0 lif \else %
 \ifnum\pdfstrcmp{#2}{f_sig}=0 \ifnum\pdfstrcmp{#1}{dec}=0 -1 \else\ifnum\pdfstrcmp{#1}{sci}=0 -1 \times 10^{0} \else\ifnum\pdfstrcmp{#1}{dec0}=0 -1 \else\ifnum\pdfstrcmp{#1}{dec1}=0 -1 \else\ifnum\pdfstrcmp{#1}{dec2}=0 -1 \else -1\fi\fi\fi\fi\fi \else %
 \ifnum\pdfstrcmp{#2}{vhist_N}=0 100 \else %
 \PackageError{paramvalue}{unknown param name: #2}{} 
 \fi\fi\fi\fi\fi\fi\fi\fi\fi\fi\fi\fi\fi\fi\fi\fi\fi\fi\fi\fi\fi\fi}
 
 \includegraphics{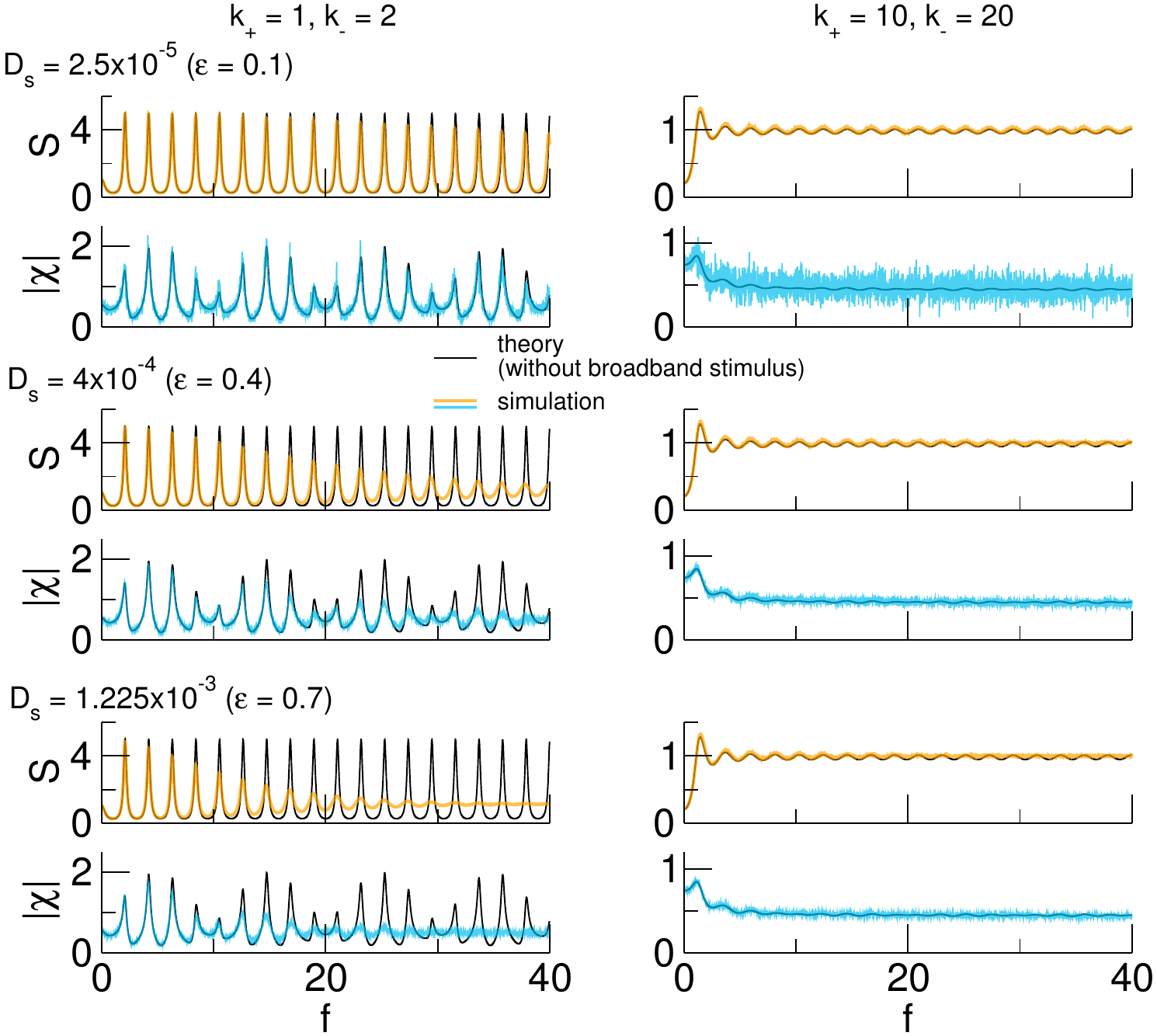}
 \caption{\label{fig:broadband} {\bf Power spectrum and susceptibility with a broadband stimulus.}  Shown are simulations for the power spectrum $S(f)$ (orange lines) and susceptibility $\chi(f)$ (blue lines), compared to the theory without broadband stimulus, \e{powspec} and \e{suscep} (black lines). Because the focus here is on the simulations with a broadband signal, we keep the lines depicting the theory in the background. We plot these quantities for two combinations of switching rates and three different values of the stimulus intensity $D_s = \eps^2/(4f_{\rm c})$, where $f_{\rm c} = \paramvalue{f_c}$ is the cutoff frequency. Remaining parameters: $\mu = \paramvalue{mu}, \s=\paramvalue{s}, v_R=\paramvalue{vr}, v_T=\paramvalue{vt}$.}
\end{figure}

The linear response ansatz, \e{linear_response_ansatz}, is in principle valid for arbitrary stimuli, as long as they are weak.
In theoretical and experimental studies \citep{Sak92}, broadband stimuli, such as band-limited Gaussian noise with a flat spectrum, have often been used, as they allow to probe the susceptibility at different frequencies simultaneously.
As we have argued, the features of power spectrum and susceptibility of neurons driven by a slowly switching dichotomous noise arise mainly because of the absence of further stochasticity within the two noise states.
A broadband stimulus acts as an additional noise source, and one can thus expect it to have a qualitative effect on these features.

{
\newcommand{\paramvalue}[2][]{\protect %
\ifnum\pdfstrcmp{#2}{f_max}=0 \ifnum\pdfstrcmp{#1}{dec}=0 100 \else\ifnum\pdfstrcmp{#1}{sci}=0 1 \times 10^{2} \else\ifnum\pdfstrcmp{#1}{dec0}=0 100 \else\ifnum\pdfstrcmp{#1}{dec1}=0 100 \else\ifnum\pdfstrcmp{#1}{dec2}=0 100 \else 100\fi\fi\fi\fi\fi \else %
\ifnum\pdfstrcmp{#2}{vhist_r}=0 \ifnum\pdfstrcmp{#1}{dec}=0 1.1 \else\ifnum\pdfstrcmp{#1}{sci}=0 1.1 \times 10^{0} \else\ifnum\pdfstrcmp{#1}{dec0}=0 1 \else\ifnum\pdfstrcmp{#1}{dec1}=0 1.1 \else\ifnum\pdfstrcmp{#1}{dec2}=0 1.1 \else 1.1\fi\fi\fi\fi\fi \else %
\ifnum\pdfstrcmp{#2}{integ_infty}=0 \ifnum\pdfstrcmp{#1}{dec}=0 15 \else\ifnum\pdfstrcmp{#1}{sci}=0 1.5 \times 10^{1} \else\ifnum\pdfstrcmp{#1}{dec0}=0 15 \else\ifnum\pdfstrcmp{#1}{dec1}=0 15 \else\ifnum\pdfstrcmp{#1}{dec2}=0 15 \else 15\fi\fi\fi\fi\fi \else %
\ifnum\pdfstrcmp{#2}{N_trials}=0 1000000 \else %
\ifnum\pdfstrcmp{#2}{df}=0 \ifnum\pdfstrcmp{#1}{dec}=0 0.01 \else\ifnum\pdfstrcmp{#1}{sci}=0 1 \times 10^{-2} \else\ifnum\pdfstrcmp{#1}{dec0}=0 0 \else\ifnum\pdfstrcmp{#1}{dec1}=0 0 \else\ifnum\pdfstrcmp{#1}{dec2}=0 0.01 \else 0.01\fi\fi\fi\fi\fi \else %
\ifnum\pdfstrcmp{#2}{vr}=0 \ifnum\pdfstrcmp{#1}{dec}=0 0 \else\ifnum\pdfstrcmp{#1}{sci}=0 0 \else\ifnum\pdfstrcmp{#1}{dec0}=0 0 \else\ifnum\pdfstrcmp{#1}{dec1}=0 0 \else\ifnum\pdfstrcmp{#1}{dec2}=0 0 \else 0\fi\fi\fi\fi\fi \else %
\ifnum\pdfstrcmp{#2}{seed}=0 0 \else %
\ifnum\pdfstrcmp{#2}{vt}=0 \ifnum\pdfstrcmp{#1}{dec}=0 1 \else\ifnum\pdfstrcmp{#1}{sci}=0 1 \times 10^{0} \else\ifnum\pdfstrcmp{#1}{dec0}=0 1 \else\ifnum\pdfstrcmp{#1}{dec1}=0 1 \else\ifnum\pdfstrcmp{#1}{dec2}=0 1 \else 1\fi\fi\fi\fi\fi \else %
\ifnum\pdfstrcmp{#2}{eps_km}=0 \ifnum\pdfstrcmp{#1}{dec}=0 0 \else\ifnum\pdfstrcmp{#1}{sci}=0 0 \else\ifnum\pdfstrcmp{#1}{dec0}=0 0 \else\ifnum\pdfstrcmp{#1}{dec1}=0 0 \else\ifnum\pdfstrcmp{#1}{dec2}=0 0 \else 0\fi\fi\fi\fi\fi \else %
\ifnum\pdfstrcmp{#2}{dt}=0 \ifnum\pdfstrcmp{#1}{dec}=0 0.005 \else\ifnum\pdfstrcmp{#1}{sci}=0 5 \times 10^{-3} \else\ifnum\pdfstrcmp{#1}{dec0}=0 0 \else\ifnum\pdfstrcmp{#1}{dec1}=0 0 \else\ifnum\pdfstrcmp{#1}{dec2}=0 0.01 \else 0.005\fi\fi\fi\fi\fi \else %
\ifnum\pdfstrcmp{#2}{vhist_l}=0 \ifnum\pdfstrcmp{#1}{dec}=0 -0.1 \else\ifnum\pdfstrcmp{#1}{sci}=0 -1 \times 10^{-1} \else\ifnum\pdfstrcmp{#1}{dec0}=0 -0 \else\ifnum\pdfstrcmp{#1}{dec1}=0 -0.1 \else\ifnum\pdfstrcmp{#1}{dec2}=0 -0.1 \else -0.1\fi\fi\fi\fi\fi \else %
\ifnum\pdfstrcmp{#2}{eps_mu}=0 \ifnum\pdfstrcmp{#1}{dec}=0 0.7 \else\ifnum\pdfstrcmp{#1}{sci}=0 7 \times 10^{-1} \else\ifnum\pdfstrcmp{#1}{dec0}=0 1 \else\ifnum\pdfstrcmp{#1}{dec1}=0 0.7 \else\ifnum\pdfstrcmp{#1}{dec2}=0 0.7 \else 0.7\fi\fi\fi\fi\fi \else %
\ifnum\pdfstrcmp{#2}{integ_epsabs}=0 \ifnum\pdfstrcmp{#1}{dec}=0 0.0001 \else\ifnum\pdfstrcmp{#1}{sci}=0 1 \times 10^{-4} \else\ifnum\pdfstrcmp{#1}{dec0}=0 0 \else\ifnum\pdfstrcmp{#1}{dec1}=0 0 \else\ifnum\pdfstrcmp{#1}{dec2}=0 0 \else 1 \times 10^{-4}\fi\fi\fi\fi\fi \else %
\ifnum\pdfstrcmp{#2}{f_c}=0 \ifnum\pdfstrcmp{#1}{dec}=0 100 \else\ifnum\pdfstrcmp{#1}{sci}=0 1 \times 10^{2} \else\ifnum\pdfstrcmp{#1}{dec0}=0 100 \else\ifnum\pdfstrcmp{#1}{dec1}=0 100 \else\ifnum\pdfstrcmp{#1}{dec2}=0 100 \else 100\fi\fi\fi\fi\fi \else %
\ifnum\pdfstrcmp{#2}{tr}=0 \ifnum\pdfstrcmp{#1}{dec}=0 0.1 \else\ifnum\pdfstrcmp{#1}{sci}=0 1 \times 10^{-1} \else\ifnum\pdfstrcmp{#1}{dec0}=0 0 \else\ifnum\pdfstrcmp{#1}{dec1}=0 0.1 \else\ifnum\pdfstrcmp{#1}{dec2}=0 0.1 \else 0.1\fi\fi\fi\fi\fi \else %
\ifnum\pdfstrcmp{#2}{km}=0 \ifnum\pdfstrcmp{#1}{dec}=0 20 \else\ifnum\pdfstrcmp{#1}{sci}=0 2 \times 10^{1} \else\ifnum\pdfstrcmp{#1}{dec0}=0 20 \else\ifnum\pdfstrcmp{#1}{dec1}=0 20 \else\ifnum\pdfstrcmp{#1}{dec2}=0 20 \else 20\fi\fi\fi\fi\fi \else %
\ifnum\pdfstrcmp{#2}{mu}=0 \ifnum\pdfstrcmp{#1}{dec}=0 0.8 \else\ifnum\pdfstrcmp{#1}{sci}=0 8 \times 10^{-1} \else\ifnum\pdfstrcmp{#1}{dec0}=0 1 \else\ifnum\pdfstrcmp{#1}{dec1}=0 0.8 \else\ifnum\pdfstrcmp{#1}{dec2}=0 0.8 \else 0.8\fi\fi\fi\fi\fi \else %
\ifnum\pdfstrcmp{#2}{s}=0 \ifnum\pdfstrcmp{#1}{dec}=0 2.4 \else\ifnum\pdfstrcmp{#1}{sci}=0 2.4 \times 10^{0} \else\ifnum\pdfstrcmp{#1}{dec0}=0 2 \else\ifnum\pdfstrcmp{#1}{dec1}=0 2.4 \else\ifnum\pdfstrcmp{#1}{dec2}=0 2.4 \else 2.4\fi\fi\fi\fi\fi \else %
\ifnum\pdfstrcmp{#2}{kp}=0 \ifnum\pdfstrcmp{#1}{dec}=0 10 \else\ifnum\pdfstrcmp{#1}{sci}=0 1 \times 10^{1} \else\ifnum\pdfstrcmp{#1}{dec0}=0 10 \else\ifnum\pdfstrcmp{#1}{dec1}=0 10 \else\ifnum\pdfstrcmp{#1}{dec2}=0 10 \else 10\fi\fi\fi\fi\fi \else %
\ifnum\pdfstrcmp{#2}{model}=0 lif \else %
\ifnum\pdfstrcmp{#2}{f_sig}=0 \ifnum\pdfstrcmp{#1}{dec}=0 -1 \else\ifnum\pdfstrcmp{#1}{sci}=0 -1 \times 10^{0} \else\ifnum\pdfstrcmp{#1}{dec0}=0 -1 \else\ifnum\pdfstrcmp{#1}{dec1}=0 -1 \else\ifnum\pdfstrcmp{#1}{dec2}=0 -1 \else -1\fi\fi\fi\fi\fi \else %
\ifnum\pdfstrcmp{#2}{vhist_N}=0 100 \else %
\PackageError{paramvalue}{unknown param name: #2}{} 
\fi\fi\fi\fi\fi\fi\fi\fi\fi\fi\fi\fi\fi\fi\fi\fi\fi\fi\fi\fi\fi\fi}

In \fig{broadband}, we plot the power spectrum and the absolute value of the susceptibility for two switching-rate combinations and three different intensities of a Gaussian stimulus with a flat spectrum of height $S_{s} = 1/(2 f_{\rm c})$, where $f_{\rm c} = \paramvalue{f_c}$ is the cutoff frequency.
The intensity of the signal is then given by $D_s = \eps^2 S_{s}/2$.
In simulations, we use a time step $\Delta t = \paramvalue{dt}$, which means that the stimulus is effectively white (the cutoff frequency corresponds to the Nyquist frequency).

In line with our reasoning above and with previous results for the power spectrum of DMP-driven PIF neurons \citep{MulDro15}, additional noise (here the broadband signal) can be seen to abolish the undamped periodicity in spectrum and susceptibility.
For small signal strength, this is hardly noticeable up to frequencies that would be considered high in neurophysiology (note that, assuming a membrane time constant $\tau_m = \unit[20]{ms}$, the dimensionless frequency $f=20$ corresponds to \unit[1000]{Hz}).
With increasing signal strength, the range where our expressions match the simulation is shifted to smaller frequencies. 
It is interesting to note that the broadband signal does not affect spectrum and susceptibility at all frequencies; instead, there seems to be a cutoff frequency, depending on the signal strength, below which the additional noise has no effect.

The influence of a broadband stimulus is more noticeable for slow switching; for faster switching, the periodicity is less prominent in the first place.
In particular for the susceptibility with fast switching, the periodicity is hardly visible and there are no qualitative differences between the different noise intensities, except for the better statistics that a stronger signal brings along.  
}

\section{Concluding remarks}
\label{sec:summary}

We have studied leaky integrate-and-fire neurons driven by an asymmetric dichotomous noise.
For this particular kind of colored noise, we were able to derive exact expressions for the spontaneous power spectrum and the susceptibility to an additive signal.
We have verified our expressions by comparison to numerical simulations.

A prominent difference to the classical results for LIF neurons driven by Gaussian white noise \citep{BruCha01,LinLSG01,LinLSG02} is the periodic structure that both the power spectrum and the susceptibility exhibit at high frequencies.
For the power spectrum, we have explained how this structure arises through delta-peaks in the spike-triggered rate. 
These stem from the fact that there is a finite probability that an ISI has a certain length (the deterministic time from reset to threshold in the plus state).
They are thus a manifestation of the discrete nature of the dichotomous noise.
As we have shown, this periodic structure of the spectrum is independent of the chosen neuron model.
The susceptibility shows a similar periodic structure. 
However, here, we lack an intuitive explanation like we found for the power spectrum.

Other differences between the DMP-driven and the Gaussian-white-noise case can be traced to the fact that the noise is colored: In contrast to LIF neurons driven by Gaussian white noise, for which the susceptibility decays to zero at a finite phase lag for increasing frequency \citep{LinLSG01,BruCha01}, here it oscillates around a finite value at zero phase lag.
This indicates that the system can respond to arbitrarily fast signals, a feature that has also been observed for LIF neurons driven by exponentially correlated Gaussian noise \citep{BruCha01} (for an in-depth discussion for more general IF models, see \cite{FouHan03}).

Especially when the noise switching is slow, the spectral measures are dominated by the periodic structure. By numerical simulations we demonstrated that this feature is robust in a certain sense against small additional fluctuations. Although  a very weak broadband Gaussian stimulus leads to a non-pathological behavior of the spectral measures, oscillatory features are still present in a wide frequency band.
Specifically, although the power spectrum of the spike train becomes flat and the susceptibility decays in the limit $f\to \infty$,  both functions still oscillate up to considerably high frequency. Remarkably, the effective cutoff frequency is set by the noise level.     

Our results are applicable to situations, where nerve cells are stimulated with a two-state input, such as a stimulus originating in a bursting neuron or input from a population of neurons that undergo up/down transitions. A further, non-obvious application is the shot-noise-limit of the Markovian dichotomous noise, which allows us to obtain expressions for an LIF neurons driven by white, excitatory Poisson input (a special case of the setup treated in ref. \citep{RicSwa10}). This will be pursued elsewhere.
\appendix

\section{}
\label{app:solsimpl}

Around $z=0$ (i.e. the stable fixed point of the minus dynamics, where $v=\mu-\s$)), linearly independent solutions to the hypergeometric equation (the homogeneous part of \e{j_ode}) are given by
\begin{align}
 \tJ_1(z) 
 		 &= \hyper{2 \pi i f, 1 - k_+ - k_- + 2 \pi i f, 1 - k_- + 2 \pi i f, z}, \\
 \tJ_2(z) 
        &= z^{k_- - 2 \pi i f} \hyper{k_-, 1-k_+;1+k_- - 2 \pi i f;z}, \label{eq:fundamentalsystem}
\end{align}
where $\hyper{a,b;c,z}$ is the hypergeometric function \citep{AbrSte72}.
In the following, we solve the inhomogeneous ODE 
\begin{equation}
 \label{eq:inhom}
 \tJ''(z)+p(z) \tJ'(z) + q(z) \tJ(z)=\til{\Delta}(z),
\end{equation}
where \begin{equation}
 \tDelta(z) = 2\s \Bigg[ \left( p(z) + \frac{2\pi i f}{1-z} \right)  \tDelta_+(z) + \left( p(z) - \frac{2\pi i f}{z} \right)\tDelta_-(z) + \tDelta_+'(z) + \tDelta_-'(z) \Bigg].
\end{equation}
The $\tDelta_\pm(u)$ are given by \es{Deltap}{Deltam}.
However, in the following we are only going to use that they vanish outside of the interval $[\min(0,z_R),z_T]$.
Given the two linearly independent solutions to the homogeneous ODE, a particular solution to \e{inhom} is known \citep{MorFes53}:
\begin{equation}
\label{eq:partsol}
 \tJ_{\rm p}(z) = \intl_{z}^{z_c} du\; \til{\Delta}(u) \frac{\tJ_2(u) \tJ_1(z) - \tJ_1(u) \tJ_2(z)}{W(u)},
\end{equation}
where $W(z)$ is the Wronskian,
\begin{equation}
 \label{eq:wronskian}
 W(z) = \tJ_1(z) \tJ_2'(z) - \tJ_1'(z) \tJ_2(z),
\end{equation}
and the upper integration limit $z_c$ can still be freely chosen.
The general solution is then given by
\begin{equation}
\label{eq:gensol}
  tJ(z) = c_1 J_1(z) + c_2 J_2(z) + tJ_{\rm p}(z).
\end{equation}

In order to fix the integration constants in \e{gensol}, one needs to distinguish two possible parameter regimes:
If $\mu-\s < v_R$, corresponding to $z_R > 0$, the fixed point in the minus dynamics, $v=\mu-s$ (corresponding to the singular point at $z=0$ in the hypergeometric differential equation) lies on the lower boundary (no trajectories can move to more negative values).
In contrast, for $\mu-\s > v_R$, corresponding to $z_R < 0$, it lies within the interval of interest.

In the first case, the integration constants in \e{gensol} can be fixed over the whole interval $[0,z_T]$ by imposing the boundary conditions at the threshold, $J(z_T^+)=0$ and $J'(z_T^+)=0$. Because $\tDelta_\pm(z>z_T) = 0$, both can be fulfilled by setting $c_1 = c_2 = 0$ and $z_c = \infty$ (equivalent to $z_c = z_T^+$, as the integrand vanishes for all $z > z_T$).
Thus, for $z_R > 0$, one has
\begin{equation}
  \tJ(z) = \intl_{z}^{\infty} du\; \til{\Delta}(u) \frac{\tJ_2(u) \tJ_1(z) - \tJ_1(u) \tJ_2(z)}{W(u)}
\end{equation}
At $z=0$, the lower boundary of the possible dynamics, the flux needs to vanish, leading to the condition 
\begin{equation}
\label{eq:zerocond}
0 = \intl_{0}^{\infty} du\; \til{\Delta}(u) \frac{\tJ_2(u) \tJ_1(0) - \tJ_1(u) \tJ_2(0)}{W(u)} = \intl_{-\infty}^{\infty} du\; \til{\Delta}(u) \frac{\tJ_2(u)}{W(u)},
\end{equation}
where we have extended the interval of integration from $[0,\infty]$ to $[-\infty,\infty]$ (the integrand vanishes for $z<0$) and used that $\tJ_1(0) = 1, \tJ_2(0) = 0$ \citep{AbrSte72}.

In the second case, $z_R < 0$, the fixed point in the minus dynamics lies within the interval in which we want to obtain a solution for $\tJ(z)$.
As pointed out previously \citep{Ben06}, one cannot expect the same solution to be valid on both sides of such a point. In particular, for the LIF neuron, this means that the integration constants need to be chosen separately in the intervals $[z_R, 0]$ and $[0, z_T]$: Above the fixed point, they need to satisfy the boundary conditions at the threshold, below it, those at the reset voltage (for details, see \citep{DroLin14,Dro15}).
The boundary conditions at the rest are satisfied by setting $z_c = -\infty$ (or any value $\le z_R^-$). One thus has
\begin{equation}
 \tJ(z) = \begin{cases}
  \intl_{z}^{-\infty} du\; \til{\Delta}(u) \frac{\tJ_2(u) \tJ_1(z) - \tJ_1(u) \tJ_2(z)}{W(u)} & z_R < z \le 0 \\
  \intl_{z}^{\infty} du\; \til{\Delta}(u) \frac{\tJ_2(u) \tJ_1(z) - \tJ_1(u) \tJ_2(z)}{W(u)} & 0 < z < z_T
  \end{cases}
\end{equation}
The condition that the flux needs to be continuous at the fixed point implies that
\begin{equation}
 \intl_{0}^{-\infty}  du\; \til{\Delta}(u) \frac{\tJ_2(u)}{W(u)} = \intl_{0}^{\infty}  du\; \til{\Delta}(u) \frac{\tJ_2(u)}{W(u)},
\end{equation}
which is equivalent to the condition that was obtained for the other case, $0 < z_R$ (\e{zerocond}). 
We can thus solve \e{zerocond} for the spike-triggered rate in both parameter regimes.
In the following, we simplify it to obtain the form given in \e{jzerosol}.

We write
\begin{align}
 \begin{split}
    0  &= \intl_{-\infty}^{\infty} du\; \Bigg[\left( p(u) + \frac{2\pi i f}{1-u} \right)  \tDelta_+(u) + \left( p(u) - \frac{2\pi i f}{u} \right)\tDelta_-(u) \\&\quad 
    + \tDelta_+'(u) + \tDelta_-'(u) \Bigg] \frac{\tJ_2(z)}{W(z)}.
  \end{split}
\end{align}
Noting that
\begin{align}
\begin{split}
 W'(z) 
       &= \tJ_1(z) \left[ -p(z) \tJ_2'(z) - q(z) \tJ_2(z) \right] 
       - \left[-p(z) \tJ_1'(z) - q(z) \tJ_1(z)\right] \tJ_2(z)
\end{split} \\
       &= -p(z) W(z),
\end{align}
the Wronskian can be written as,
\begin{equation}
 \label{eq:wronskisol}
 W(z) = c_W \cdot e^{-\int^z dw\; p(w)} = c_W \cdot (1-z)^{k_+-1-2\pi i f} \cdot z^{k_- - 1- 2\pi i f},
\end{equation}
where $c_W$ is a constant that will drop out. 
Further,
\begin{equation}
 \frac{\partial}{\partial z} \left[ \frac{\tJ_2(z)}{W(z)} \right] = \frac{\tJ_2'(z)+p(z)\tJ_2(z)}{W(z)},
\end{equation}
Integrating by parts and exploiting that $\tDelta_\pm(z)$ vanishes at the boundaries yields, 
\begin{align}
\begin{split}
 0 &= \intl_{-\infty}^{\infty} du\; \frac{\tDelta_+(u)}{W(u)} \left( \frac{2\pi i f}{1-u} \tJ_2(u)-\tJ_2'(u) \right)  - \frac{\tDelta_-(u)}{W(u)} \left( \frac{2\pi i f}{u} \tJ_2(u)+\tJ_2'(u) \right).
 \end{split}
\end{align}
This can be further simplified using known properties of hypergeometric functions:
Using \citein[15.2.4]{AbrSte72},
\begin{align}
 \begin{split}
 \tJ_2'(z) &= \left[ z^{k_- - 2\pi i f} \cdot \hyper{k_-, 1-k_+;1+k_- - 2\pi i f;z} \right]' \\
  &= (k_- - 2\pi i f) \cdot z^{k_- - 2\pi i f -1} \cdot \hyper{k_-, 1-k_+;k_- - 2\pi i f;z}
 \end{split}
\end{align}
and thus, via \citein[15.2.25]{AbrSte72}
\begin{align}
 \begin{split}
  \frac{2\pi i f}{1-z} \tJ_2(z) - \tJ_2'(z) &= -z^{k_- - 2\pi i f - 1} (1-z)^{-1} (k_- - 2\pi i f) \cdot \hyper{k_-,-k_+;k_- - 2\pi i f;z}, \\
 \end{split}
\end{align}
and (\citein[15.2.17]{AbrSte72})
\begin{align}
 \frac{2\pi i f}{z} \tJ_2(z) + \tJ_2'(z) &= z^{k_- - 2\pi i f - 1}  k_- \cdot \hyper{1+k_-,1-k_+;1-k_- - 2\pi i f;z}.
\end{align}
Plugging in \e{wronskisol} for the Wronskian and using \citein[15.3.3]{AbrSte72}, one finds
\begin{align}
 \frac{1}{W(z)}& \left( \frac{2\pi i f}{1-z} \tJ_2(z) - \tJ_2'(z) \right) \\
  &= - c_W^{-1} (k_- - 2\pi i f) \cdot \hyper{-2\pi i f,k_+ + k_- - 2\pi i f;k_- - 2\pi i f;z}
   \\ &= - c_W^{-1} (k_- - 2\pi i f)  \mathcal{F}(z, f),
\end{align}
and
\begin{align}
 \frac{1}{W(z)}& \left( \frac{2\pi i f}{z} \tJ_2(z) + \tJ_2'(z) \right) \\
  &=  c_W^{-1} k_- \cdot \hyper{-2\pi i f,k_+ + k_- - 2\pi i f;1+k_- - 2\pi i f;z}
   \\ &=  c_W^{-1} k_-  \mathcal{G}(z, f),
\end{align}
and thus arrives at the expression given in \e{jzerosol}.

\section{}
\label{app:highfreq_equiv}

Here, we show the equivalence of \e{powspec_highfreq} and \e{s_delta}.
We start from \e{s_delta}. Introducing the abbreviations
\begin{equation}
 a := \frac{\hat{k}_+ T_d^+}{2\pi}, \quad b:= f T_d^+,
\end{equation} it takes the form
\begin{align}
S_\delta(f) &= \frac{r_0}{T_d^+} \suml_{n=-\infty}^{\infty} \frac{2 \hat{k}_+}{\hat{k}_+^2 + \left(2 \pi \left[f - \frac{n}{T_d^+} \right]\right)^2} \\
&= \frac{r_0}{2\pi} \suml_{n=-\infty}^{\infty} \frac{2a}{a^2+(b-n)^2} 
\\ &= \frac{r_0}{2\pi} \left[ \frac{2 a}{a^2+b^2} + \suml_{n=1}^{\infty} \frac{1}{a+i (b-n)} + \frac{1}{a+i (b+n)} + \frac{1}{a-i (b-n)} + \frac{1}{a+i (b+n)}\right] 
\\&= \frac{r_0}{2\pi} \left[ \frac{2 a}{a^2+b^2} + \suml_{n=1}^{\infty} \frac{2 (a+ ib)}{(a+ib)^2+n^2} + \frac{2 (a- ib)}{(a-ib)^2+n^2} \right].
\end{align}
Using \citein[1.421.3]{GraRyz65}
\begin{equation}
 \coth\left(\pi x\right) = \frac{1}{\pi x} + \frac{2x}{\pi} \suml_{n=1}^{\infty} \frac{1}{x^2+n^2},
\end{equation}
we have
\begin{align}
 S_\delta(f)
 &= \frac{r_0}{2} \left[\coth\left(\pi \left[a + ib\right]\right) + \coth\left(\pi \left[a - ib\right]\right)\right].
\end{align}
Using $\coth x = (\exp[x]+\exp[-x])/(\exp[x]-\exp[-x])$ and reinserting $a$ and $b$, it is straightforward to show that this yields
\begin{align}
 S_\delta(f) &= r_0 \frac{\sinh\left(\hat{k}_+ T_d^+\right)}{\cosh\left(\hat{k}_+ T_d^+\right) - \cos\left(2\pi f T_d^+\right)} = S(f \gg 1),
 \end{align}
i.e. the compact expression for the high-frequency limit, \e{powspec_highfreq_compact}.

\section{}
\label{app:susdetails}

It is convenient to transform the dynamics \e{dynamics_lif} by using
\begin{equation}
 x := v + \varepsilon \frac{1}{2 \pi i f-1} e^{-2 \pi i f t}.
\end{equation}
This yields a system without additive signal,
\begin{equation}
 \dot x = \mu - x +  \eta(t),
\end{equation}
at the cost of introducing time-dependent reset and threshold values,
\begin{equation}
 x_R(t) = v_R  + \varepsilon \frac{1}{2 \pi i f-1} e^{-2 \pi i f t}, \quad x_T(t) = v_T  + \varepsilon \frac{1}{2 \pi i f-1} e^{-2 \pi i f t}.
\end{equation}
The new master equations read
\begin{align}
 \begin{split}
 \label{eq:dichotheo:suscep_master_x_1}
 \partial_t P_+(x,t) &= - \partial_x \left( (\mu-x+\s) P_+(x,t) \right) - k_+ P_+(x,t) + k_- P_-(x,t) \\
 &\quad + r(t-\tr) P_{+|+}(\tr) \delta[x - x_R(t)] - r(t) \delta[x - x_T(t)],
 \end{split} \\
 \label{eq:dichotheo:suscep_master_x_2}
 \begin{split} 
 \partial_t P_-(x,t) &= - \partial_x \left( (\mu-x-\s) P_-(x,t) \right) + k_+ P_+(x,t) - k_- P_-(x,t) \\
  &\quad + r(t-\tr) P_{-|+}(\tr) \delta[x - x_R(t)],
 \end{split}
\end{align}
with the same trivial boundary conditions as above.
Plugging in \e{r_ansatz} and \e{p_ansatz}, Taylor-expanding the $\delta$ functions for small $\varepsilon$, and keeping only the linear order in $\varepsilon$ yields
\begin{align}
 \begin{split}
 -2 \pi i f  P_{+,1}(x) &= - \partial_x \left( (\mu-x+\s) P_{+,1}(x) \right) - k_+ P_{+,1}(x) + k_- P_{-,1}(x) \\
 &\quad + \chi(f) \left[e^{2 \pi i f\tr}  P_{+|+}(\tr) \delta(x - v_R) - \delta(x - v_T ) \right] \\
 &\quad - \frac{r_0}{2 \pi i f - 1} \left[ P_{+|+}(\tr) \delta'(x - v_R) - \delta'(x - v_T) \right],
 \end{split} \\
 \begin{split}
  -2 \pi i f P_{-,1}(x) &= - \partial_x \left( (\mu-x-\s) P_{-,1}(x) \right) + k_+ P_{+,1}(x) - k_- P_{-,1}(x) \\
    &\quad + \chi(f) e^{2 \pi i f\tr}  P_{-|+}(\tr) \delta(x - v_R)\\
    &\quad - \frac{r_0}{2 \pi i f - 1} P_{-|+}(\tr) \delta'(x - v_R).
  \end{split}
\end{align}
This has the same structure as the Fourier-transformed version of \es{strate_master1}{strate_master2}.
The correction to the flux, $J_1(x)$, with
\begin{equation}
 J(x,t) = J_0(x) + \varepsilon e^{-2 \pi i f t} J_1(x),
\end{equation}
then follows the ODE \e{j_ode}, if the inhomogeneities $\tDelta_\pm(x)$ are appropriately chosen, and \e{jzerosol} can be used to extract $\chi(f)$.
The derivatives of $\mathcal{F}(z,f)$ and $\mathcal{G}(z,f)$, which appear when integrating by parts, are given by \citep{AbrSte72}
\begin{align}
\begin{split}
 \mathcal{F}'(z,f) &= \frac{-2 \pi i f(k_+ +k_--2 \pi i f)}{k_- - 2 \pi i f} \hyper{1-2 \pi i f,1+ k_+ + k_- - 2 \pi i f;1+k_- - 2 \pi i f;z}, \\
\end{split}\\
\begin{split}
 \mathcal{G}'(z,f) &= \frac{-2 \pi i f(k_++k_--2 \pi i f)}{1+k_- - 2 \pi i f} \hyper{1-2 \pi i f,1+ k_+ + k_- - 2 \pi i f;2+k_- - 2 \pi i f;z}.\\
\end{split}  
\end{align}

\end{document}